\documentclass[a4paper,11pt]{article}
\pdfoutput=1 

\usepackage{jcappub} 

\usepackage[T1]{fontenc} 
\usepackage[normalem]{ulem}
\usepackage{float}

\title{\boldmath Resonant Scattering between Dark Matter and Baryons:  Revised Direct Detection and CMB Limits}

\author{Xingchen Xu}
\author{and Glennys R. Farrar}
\affiliation{Center for Cosmology and Particle Physics, Department of Physics, New York University\\New York, NY 10003, USA}

\emailAdd{xingchen.xu@nyu.edu}
\emailAdd{glennys.farrar@nyu.edu}

\abstract{Traditional dark matter models, eg. WIMPs, assume dark matter is weakly coupled to the standard model so that elastic scattering between dark matter and baryons can be described perturbatively by Born approximation. Most direct detection experiments are analyzed according to that assumption. 
We show that when the fundamental DM-baryon interaction is attractive, dark matter-nucleus scattering is non-perturbative in much of the relevant parameter range.  The cross section exhibits rich resonant behavior with a highly non-trivial dependence on atomic mass; furthermore, the extended rather than point-like nature of nuclei significantly impacts the cross sections and must therefore be properly taken into account. The repulsive case also shows significant departures from perturbative predictions and also requires full numerical calculation.  These non-perturbative effects change the boundaries of exclusion regions from existing direct detection, astrophysical and CMB constraints.  

Near a resonance value of the parameters the typical velocity-independent Yukawa behavior, $\sigma \sim v^{0}$, does not apply; we take the non-trivial velocity dependence into account, however this more accurate treatment has little impact on limits given current constraints.  Correctly treating the extended size of the nucleus and doing an exact integration of the Schroedinger equation does have a major impact relative to past analyses based on Born approximation and naive form factors, so is essential for interpreting observational constraints.  

We report the corrected exclusion regions superseding previous limits from XQC, CRESST Surface Run, CMB power spectrum and extensions with Lyman-$\alpha$ and Milky Way satellites, and Milky Way gas clouds. Some limits become weaker, by an order of magnitude or more, than previous bounds in the literature which were based on perturbation theory and point-like sources, while others become stronger.  Gaps which open by correct treatment of some particular constraint can sometimes be closed using a different constraint.    We also discuss the dependence on mediator mass and give approximate expressions for the velocity dependence near a resonance.   
Sexaquark ($uuddss$) DM with mass around 2 GeV, which exchanges QCD mesons with baryons, remains unconstrained for most of the parameter space of interest.

A statement in the literature that a DM-nucleus cross section larger than $10^{-25}\,{\rm cm}^2$ implies dark matter is composite, is corrected.}

\keywords{Dark Matter, Direct Detection, Non-perturbative, HIDM, sexaquark}

\begin{document} 
\maketitle
\flushbottom

\newpage

\section{\label{sec:introduction}Introduction}

A possible non-gravitational interaction between dark matter and standard model particles is important both theoretically and experimentally. Such an interaction, if it exists, will have direct consequences for cosmology, astrophysics and direct detection experiments. In general, the elastic scattering of DM by baryons with a massive mediator, $\phi$, can be described by a Yukawa potential in the non relativistic limit
\begin{equation}
\label{eq:yukawa}
V(r)=-\frac{\alpha}{r}e^{-m_\phi r},
\end{equation}
in which the minus sign is for convenience so that $\alpha>0$ corresponds to an attractive potential. (In Sec.~\ref{sec:repulsive} and associated figures dealing with the repulsive case, we use (\ref{eq:yukawa}) without the minus sign to keep $\alpha$ positive.) In this paper we devote greatest attention to the attractive case because of its rich and sometimes dramatic phenomenology; the repulsive case is treated as well.

Among many dark matter models, the recently proposed model of \emph{ sexaquark}~\cite{Farrar:2017eqq,fudsDM18,Farrar:2020} dark matter (SDM) is worth special consideration by virtue of its economy and predictive power.  Our analysis is however general and applicable to Beyond the Standard Model scenarios for DM mass in the 0.1 - 100 GeV range. In the SDM model, the dark matter is a scalar particle consisting of six standard model quarks ($uuddss$), with low enough mass that its lifetime is sufficiently greater than the age of the Universe.   
The upper limit on mass is $m_S < m_\Lambda + m_p + m_e \approx 2.05$ GeV, to ensure that its decay is doubly-weak and the lifetime is longer than the age of the universe, while a mass less than $\approx 1.7$ GeV would be very difficult to reconcile with deuteron stability. We adopt $m_S \sim 2 m_p$ as a fiducial mass choice. The interaction between sexaquark and baryon is mediated by the flavor singlet combination of $\omega-\phi$ mesons. Thus the mediator mass $m_\phi$ is around the GeV scale for the sexaquark model, while for a hidden sector DM model $m_\phi$ could be quite different. 

The coupling strength $\alpha$ can naturally be as large as $\mathcal{O}(1)$, typical for a strong interaction process, although it could be much smaller depending on how $\phi$ couples to the DM. We will focus on the spin-independent (SI) cross section for simplicity; for scalar DM including sexaquark DM, this is the only case.  

Exact analytic solutions for the Yukawa potential scattering problem do not exist, and Born approximation does not apply for the parameter space we are interested in.  Moreover the extended nature of the nucleon or nucleus sourcing the Yukawa potential means that the overall potential seen by the DM is not a Yukawa, even for a proton target.  Therefore a full numerical solution is necessary. As we will see below, the typical cross section for the sexaquark model is around mb or $10^{-27}\text{\,cm}^{2}$, if the DM and mediator masses are around a GeV and $\alpha \sim \mathcal{O}(1)$. Such a large cross section can be constrained by cosmological limits~\cite{Dvorkin:2013cea,Gluscevic:2017ywp,Xu:2018efh}
and surface detectors~\cite{Erickcek:2007jv,Mahdawi:2017cxz,Mahdawi:2018euy}, while deep underground detectors are mostly insensitive due to a thick overburden shading the DM flux~\cite{Starkman:1990nj}. 

\begin{figure}
\centering 
\includegraphics[width=0.7\textwidth]{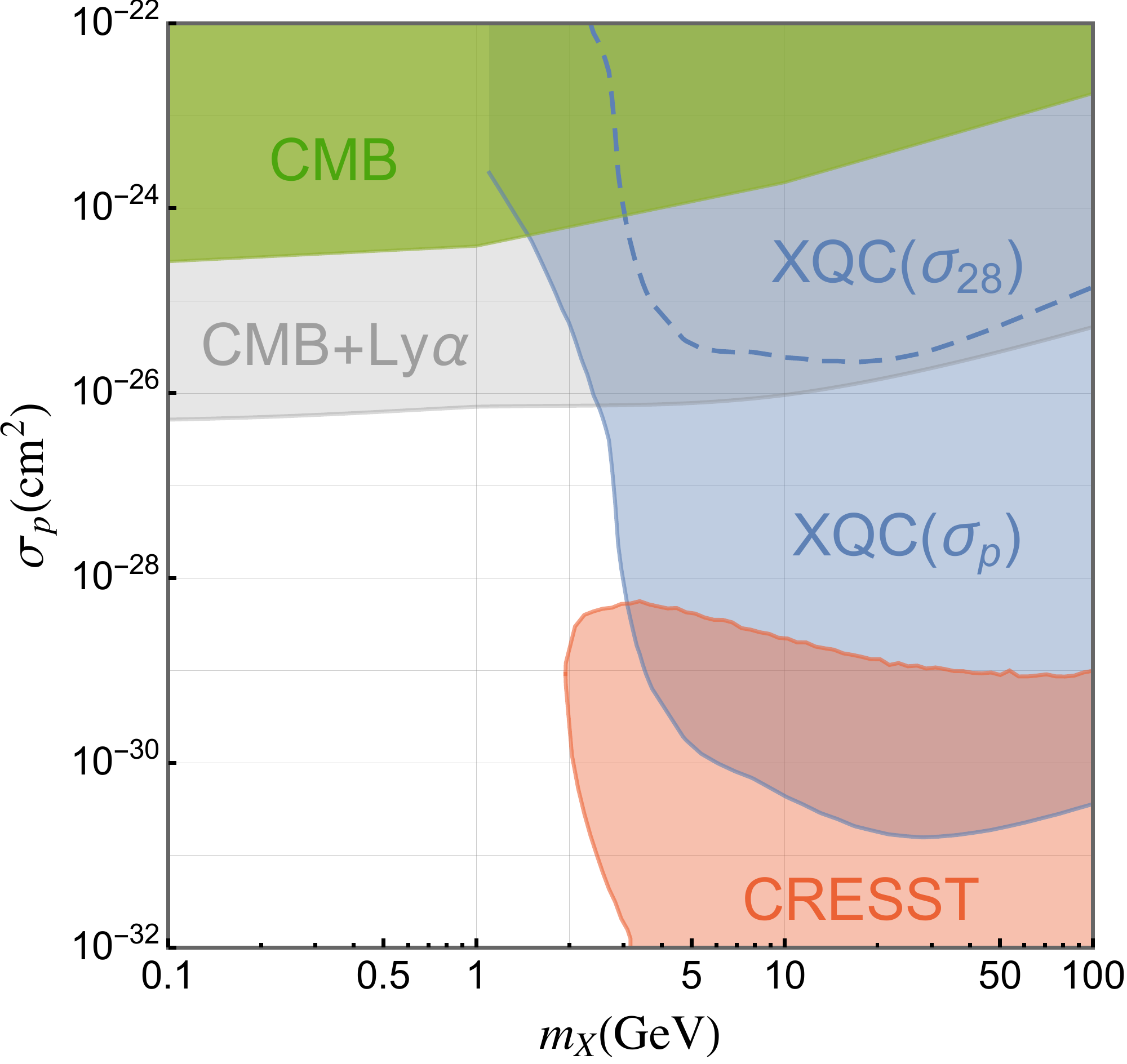}
\caption{\label{fig:experiments} Born-approximation-based DM-nucleon spin-independent cross section limits from XQC~\cite{Mahdawi:2018euy} (blue); CMB (green) and CMB+Lyman-$\alpha$ (gray) ~\cite{Xu:2018efh}; and CRESST surface detector (red)~\cite{Angloher:2017sxg,Mahdawi:2018euy}. The dashed line is the limit on the DM-Si cross section, $\sigma_{28}$, which is the actual constraint from XQC. For XQC and CRESST we display the limit corresponding to a thermalization efficiency of 0.01; simple estimates suggest a higher efficiency is implausible~\cite{Mahdawi:2018euy,Farrar:2020}.  A measurement of the thermalization efficiency is critically needed.  To reiterate:  the limits shown in this plot are generally invalid and must be replaced by those in Fig.~\ref{fig:sigmxAll} obtained using a fully non-perturbative treatment. }
\end{figure}

Figure~\ref{fig:experiments} shows the main current limits on the cross section as derived assuming Born approximation.  However this parameter region is actually largely in the non-perturbative regime, which has several consequences. The cross section has a non-trivial dependence on all of the parameters ($m_X,m_A,r_A,m_\phi,\alpha$), where $\{m_A,r_A\}$ are the target nucleus mass and radius and $A$ is the atomic mass number.  Therefore, a complete, {\it ab initio} exclusion analysis on the parameter space is needed. In particular, the cross section and atomic mass $A$ do not have a simple, model-independent scaling relationship.\footnote{We presented our results on the generic failure of the Born approximation scaling relation for DM scattering on nuclei through a Yukawa potential in~\cite{Xu:2019}.  Subsequently,  Ref.~\cite{Digman:2019} considered a repulsive square-well potential and contact interaction for massive dark matter, showing that Born approximation applies for DM-proton scattering only if $\sigma_p \ll 10^{-25}\rm{\,cm^2}$, and that for Xenon (A=131) Born scaling fails when $\sigma_p \gtrsim 10^{-32}\rm{\,cm^2}$. See Sec.~\ref{sec:digman} regarding an inaccurate assertion of~\cite{Digman:2019}.}  Without Born approximation, it is non-trivial to translate from the actual experiment measuring the interaction between DM and detector target (e.g., $A$=28 for Si in XQC) to the DM-nucleon cross section ($A$=1).   Doing so for some of the key constraining experiments and observations is the purpose of this paper.

In our analysis for the attractive interaction, resonant scattering plays an important role. We emphasize here that the resonance we are talking about in this paper should not be confused with the Breit-Wigner resonance typically seen in high energy physics. A Breit-Wigner resonance generally appears as a peak at certain center of mass energy in the function $\sigma(v)$ or $\sigma(E)$ and is usually associated with excitation of some intermediate state, see Ref~.\cite{Bai:2009cd} for example. On the other hand, the resonance we encounter here is the low energy elastic scattering s-wave resonance, which corresponds to a zero energy bound state of the scattering potential. As a result the resonance appears as a peak of the cross section at some particular parameter values ($\alpha$,$m_\phi$,$m_X$,$A$), but not velocity. For low energy scattering, $\sigma \sim v^{-2} $ on the resonance while $\sigma \sim v^0 $ off the resonance. The fact that dark matter has a velocity distribution does not smooth out the resonance as a function of the model parameters, contrary to what is perceived in~\cite{Digman:2019}.
For certain parameter choices it is possible to have a p-wave or higher wave Breit-Wigner resonance in $\sigma(v)$, which is associated with a quasi-bound state of the effective potential including the angular term. Such higher partial-wave resonances are less relevant for us as they require higher energy and are usually subdominant to the s-wave contribution.  The transition to the classical regime entails an arbitrarily large number of partial waves.
If one wants to make an analogy, the non-perturbative cross section and s-wave resonance here are closely related to the Sommerfield enhancement of DM annihilation and freeze out~\cite{Sommerfeld:1931qaf,Nima:2009th,Agrawal:2020lea}. In fact, resumming all box diagrams responsible for the Sommerfield enhancement has been shown to recover the numerical solution of the Schrödinger equation in the appropriate regime.  

Although motivated in large part by the sexaquark, our results are more general and applicable to any interaction described by a Yukawa potential in the non-relativistic quantum regime, sourced by an extended nuclear distribution.  The experimental constraints obtained here can be directly applied to any DM model with mediator mass above a few hundred MeV, using a scaling law we derive.  For lower mediator mass the methodology is applicable but the numerical experimental limits need to be re-calculated as we do for several illustrative cases. Our results can be applied to interactions within a complex hidden sector as well. We take the dark matter particle to be pointlike, but our techniques are applicable to extended dark matter case and general features of our results apply there as well.

This paper is organized as follows. In section \ref{sec:YukExt} we show some general results on non-perturbative effects and their application to experimental results. In sections~\ref{sec:reinterpret} and~\ref{sec:CMBast} we present how to reinterpret the result of direct detection experiments and astrophysical and cosmological constraints in this non-perturbative regime. We give  the combined constraints on parameter space in section~\ref{sec:attractiveresults} for attractive interaction and section~\ref{sec:repulsive} for repulsive, discuss dependence on mediator mass in section~\ref{sec:mphi} and conclude in section~\ref{sec:conclusion}.  Our numerical methods for calculating the DM-baryon scattering cross sections are described in the Appendix, where an approximate expression for the velocity dependence near a resonance is also derived.

\section{\label{sec:YukExt} Yukawa Interaction with Extended Source}

\subsection{Source Model}
\label{sec:ext}
In a realistic model where a nucleus is the source for the potential scattering of a DM particle, the source has a specified matter distribution rather than being a singular point as in the simple Yukawa of eq.~\eqref{eq:yukawa}. The Yukawa charge is then smeared out with some charge density and the potential becomes
\begin{equation}
\label{eq:potential_int}
V(\vec{r})=\int-\frac{\alpha \rho(\vec{r'})}{|\vec{r}-\vec{r'}|}e^{-m_\phi |\vec{r}-\vec{r'}|}d^3r'~,
\end{equation}
where $\rho(\vec{r'})$ is the normalized Yukawa-charge distribution of the source with normalization 
\begin{equation}
\label{eq:sourcenorm}
\int \rho(\vec{r'})d^3r'=1~.
\end{equation}
For example a point source has $\rho(\vec{r'})=\delta(\vec{r'})$. The finite size of the source regulates the Yukawa potential at the origin and thus influences the cross section.

We adopt in this paper a simple model for the nuclear density distribution as being a uniform ball with radius $r_0$, which we identify as the radius of the nucleus:
\begin{equation}
\label{eq:rhoball}
\rho(\vec{r'})=
\begin{cases}
\frac{3}{4\pi r_0^3} \quad &(r' < r_0)
\\
0 \quad &(r'\geq r_0)  
\end{cases}~~.
\end{equation}
The corresponding rms radius is 0.77 $r_0$, so given that the proton charge radius is 0.8 fm, we take $r_0=1$ fm for the proton.  We take $r_0 = R_0 A^\frac{1}{3} \equiv $ fm for nucleus of mass number $A$ with $R_0=1.0$ fm. This is a common description of the nucleus and was adopted previously for the binding of sexaquark with nuclei~\cite{Farrar:2003gh}.  It suffices for displaying the features of the extended distribution, which is our aim in this paper. We check the sensitivity of results to $R_0$ by also calculating for $R_0 = 1.2$ fm and find that the detailed position of resonance and anti-resonance features are sensitive to $R_0$ and the profile of the density distribution, so limits would vary somewhat if a smoother nuclear wave function or different $R_0$ were adopted.   

Integrating eq.~\eqref{eq:potential_int} to get the potential, we find


\begin{equation}
\label{eq:Vball}
V(r)=-\frac{3 \alpha}{m_{\phi}^{2} r_0^3} \times
\begin{cases}
1-(1+m_\phi r_0) e^{-m_\phi r_0}\frac{\sinh{(m_\phi r)}}{m_\phi r}&\!\!\!\!\! \quad (r < r_0)
\\
\left[ m_\phi r_0 \cosh{(m_\phi r_0)} - \sinh{(m_\phi r_0)}  \right] \frac{e^{-m_{\phi} r}}{m_\phi r}&\!\!\!\!\! \quad (r\geq r_0)~.
\end{cases}
\end{equation}
Now, a new length scale $r_0$ has been introduced in addition to the Yukawa screening length $\lambda=1/m_\phi$. In the limit $r_0 \ll \lambda$ we recover the point Yukawa potential~\eqref{eq:yukawa}. When $r_0 \gg \lambda$, inside the ball,  the potential is essentially constant:
\begin{equation}
\label{eq:Vballinside}
V(r)\xrightarrow[]{r_0 \gg \lambda} -\frac{3 \alpha}{m_{\phi}^{2} r_0^3} \quad (r < r_0)~~.
\end{equation}
So the potential is a square well with radius $r_0$, with a soft transition region at the boundary whose width is $\lambda$. The range of the potential is now primarily determined by $r_0$ instead of $\lambda$, when $r_0 \gg \lambda$ -- as is the case for $\lambda^{-1} \approx$ GeV as relevant for flavor singlet hadronic interactions, especially for heavy nuclei.

\subsection{General Results in the Non-perturbative Regime}
\label{sec:GenResNonPert}
As discussed in the Appendix, it is useful to write everything in dimensionless language.  For the simple Yukawa of eq.~\eqref{eq:yukawa} and in the non-relativistic limit, two parameters suffice~\cite{Buckley:2010}:

\begin{equation}
\label{eq:ab}
a \equiv \frac{v}{2\alpha} \,,
\quad\quad\quad\quad
b \equiv \frac{2 \mu \alpha}{m_\phi} \, ,
\end{equation}
with $\mu$ the reduced mass and $v$ the relative velocity.  To describe the extended nucleus we introduce a third dimensionless parameter $c$, with
\begin{equation}
\label{eq:c}
\frac{c}{b}  \equiv \frac{r_0}{\lambda} = m_\phi r_0~,
\end{equation}
so the dimensionless potential $\tilde{V}(x)$ in the Schrödinger equation \eqref{eq:schdimless} becomes, with $x \equiv 2 \mu \alpha r$: 
\begin{equation}
\label{eq:vdimlessball}
\tilde{V}(x)=-3\left(\frac{b}{c}\right)^3 \times
\begin{cases}
\frac{1}{b}-(1+\frac{c}{b}) e^{-\frac{c}{b}}\frac{1}{x} \sinh{(\frac{x}{b})} \quad &(x < c)
\\
\left[ \frac{c}{b} \cosh{(\frac{c}{b})} - \sinh{(\frac{c}{b})}  \right] \frac{1}{x} e^{-\frac{x}{b}} \quad &(x\geq c)
\end{cases}~~.
\end{equation}
We then use the methods described in the Appendix to solve for the cross section. 

\begin{figure}
\centering 
\includegraphics[width=.8\textwidth]{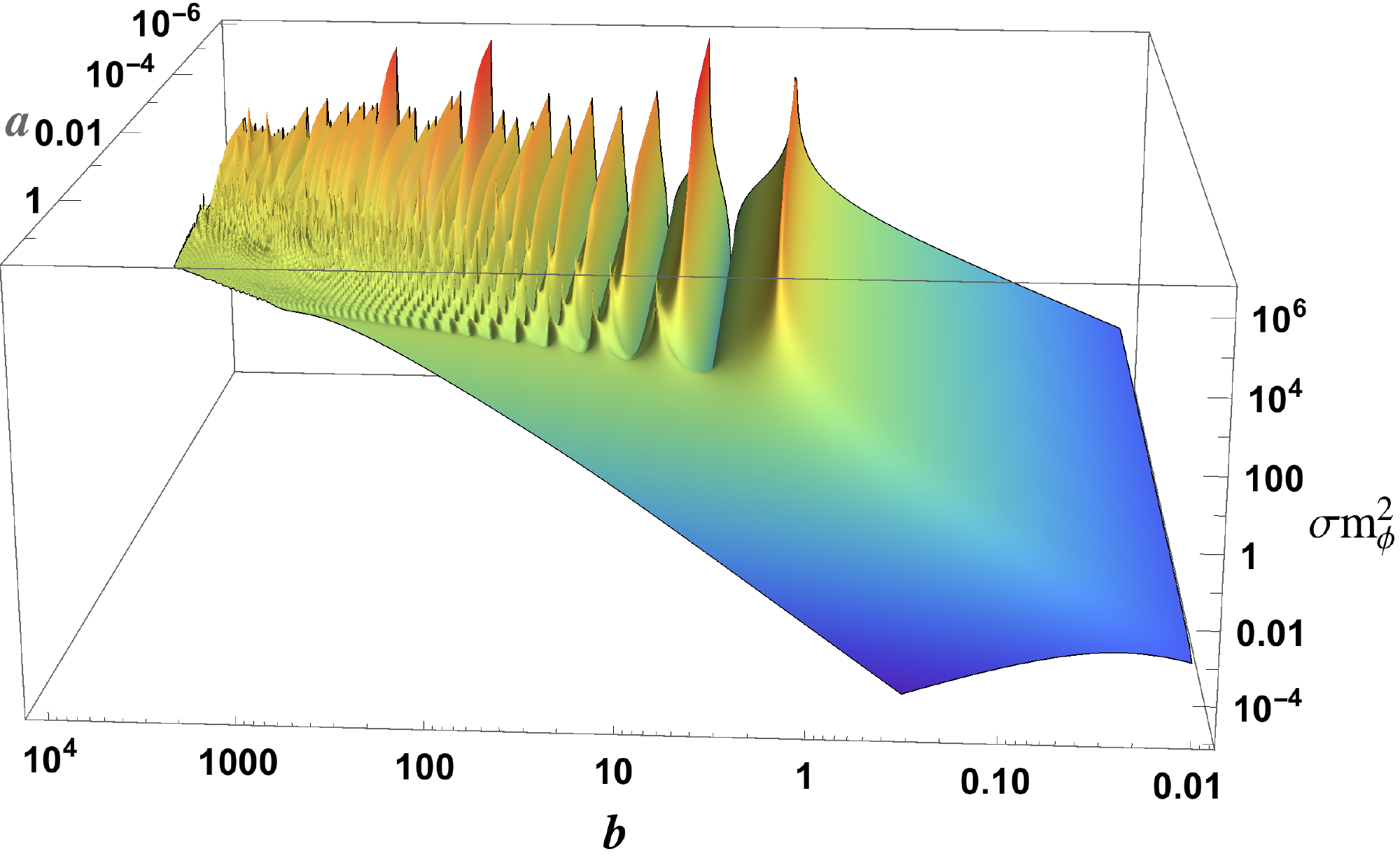}
\caption{\label{fig:3Dab} 3D plot of $\sigma m_\phi^2$ as a function of $(a,b)$ for a point source, with $\sigma m_\phi^2$ along the z-axis.  The "crenellated" appearance is merely a sampling artifact due to the resolution of the figure.}
\end{figure}

Figures~\ref{fig:2Dab} and~\ref{fig:3Dab} show $\sigma m_\phi^2$ as a function of the dimensionless parameters $(a,b)$ for the point source potential. It is obvious that the cross section is not a smooth function of the underlying parameters. For the extended potential the behaviors are similar, with shifted locations of the resonances and anti-resonances.  Fig.~\ref{fig:ComparePotential} in the Appendix shows the profile for several values of $c/b$ in comparison to a square-well.

\begin{figure}
\centering 
\includegraphics[width=.8\textwidth]{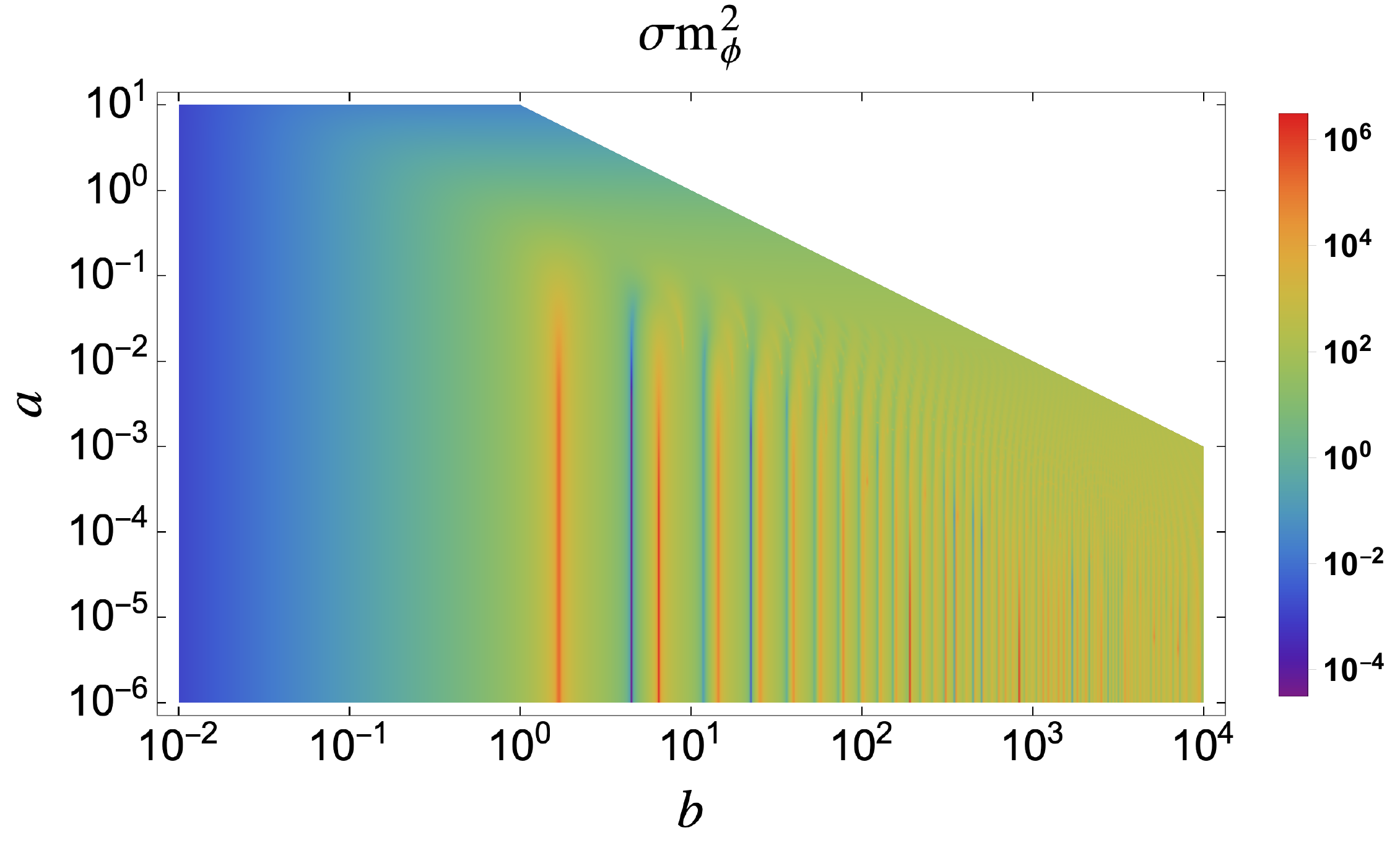}
\caption{\label{fig:2Dab} 2D plot of $\sigma m_\phi^2$ as a function of $(a,b)$ for a point source, with $\sigma m_\phi^2$ shown by colors. The cut-off in the upper right corner is where $ab\geq10$, which is the classical regime.}
\end{figure} 

\paragraph{Distinct Regimes:}  $ $\\
\underline{\textit{Classical Regime}} ($ab \gg 1$): The upper right corner in Fig.~\ref{fig:2Dab} (which is the lower left corner in Fig.~\ref{fig:3Dab}) is cut-out beyond $ab=\mu v / m_\phi \geq 10$. In this regime the De Broglie wavelength of the particle is much smaller than the typical scale of the potential and classical mechanics is sufficient to describe the scattering. In this regime there are already some non-perturbative effects relevant in plasma physics where a screened Coulomb potential is used~\cite{Khrapak:2003kjw}. Ref.~\cite{Loeb:2010gj} considered this regime in trying to solve astrophysical problems with self-interacting dark matter (SIDM) models.  Outside this region, when $ab \lesssim 1$, a classical treatment fails and quantum effects need to be included. \\\underline{\textit{Born Regime}} ($b \ll 1$): The blue region with relatively small cross section and $b \ll 1$ is the Born regime where perturbative calculation is reliable.\\ \underline{\textit{Quantum Resonant Regime}} ($ab \lesssim 1 \text{ and } b \gtrsim 1$): The region between the classical and Born regime is the quantum resonant regime where the scattering problem is entirely quantum mechanical and non-perturbative. The resonant behavior of the cross section is clear in Fig. \ref{fig:2Dab} and Fig.~\ref{fig:3Dab}. We are mostly interested in this regime where analytic description is absent and numerical calculation is a must. One model which falls in this regime is sexaquark dark matter, which has $m_X \sim 2\text{\,GeV}$ and a mediator $m_\phi \sim 1\text{\,GeV}$ with coupling possibly as large as $\alpha \sim 1$ to nucleon. For a sexaquark colliding with silicon ($m_{A} \approx 28 m_p \approx 26.3 \text{\,GeV}$) at $v=300\, \text{km/s}$, the corresponding parameters are $a=0.0005$, $b=3.71$ and $ab=0.0019$.

For the extended potential we can almost draw the same conclusions, except that when $c \gg b$, the range of the potential is determined by $r_0$ rather than $1/m_\phi$, and the quantum resonant regime is determined by $ac \ll 1$. Taking the same sexaquark-silicon collision as an example and using $ \sim A^{\frac{1}{3}}\, \text{\,fm}$ to approximate the silicon nucleus radius, we find $c/b \sim 15$ and $ac \sim 0.03$, which falls well in the quantum resonant regime.

\paragraph{Resonance and Anti-Resonance:}
The strong enhancement or diminution of cross section in the resonant regime can be understood from the phase shift. In the parameter regime that Born approximation is valid, the cross section is
\begin{equation}
\label{eq:sm-cxtotborn}
\sigma^{\rm{Born}} =\frac{4\pi b^2}{m_\phi^2 (1+4 a^2 b^2)}~,
\end{equation}
but this cannot be applied to the resonant regime.
The general result at low energy when s-wave ($l=0$) scattering is dominant (which is usually true for us, as we will see later) is,  from eq.~\eqref{eq:cxtotSI}:
\begin{equation}
\label{eq:cxtotSwave}
\sigma_\text{s-wave}=\frac{4\pi}{a^2 b^2 m_\phi^2}\sin^2(\delta_{0})~,
\end{equation}
where $\delta_{0}$ depends on $a,\, b, \, c$; $~\delta_{0}$ must be calculated numerically in the resonant regime. 
When $\delta_{0} \rightarrow \frac{\pi}{2}$ the cross section is on resonance and reaches its maximum value, resulting in the peaks in Fig.~\ref{fig:2Dab} and Fig.~\ref{fig:3Dab}. The position of the peaks is in one-to-one correspondence with the zero energy bound states of the Yukawa potential. In the pointlike-source problem ($c=0$), the potential is $e^{-\frac{x}{b}}$, with $b$ setting its range. When $b \ll 1$ the potential is too narrow and weak to accommodate any bound states. As $b$ increases, the potential becomes wider and stronger up to the point where a bound state with $E_0 \rightarrow 0^-$ emerges, in which case the scattering cross section reaches its maximum. As $b$ continues to increase, the ground state binding energy gets more and more negative, up to some point where another bound state with $E_1 \rightarrow 0^-$ emerges and the scattering cross section hits another peak. The position of these zero energy bound states are easily calculated to be at $b=1.68, 6.45, 14.34$ etc., which are exactly the locations of the peaks in Fig.~\ref{fig:2Dab} and Fig.~\ref{fig:3Dab}. On the other hand when $\delta_{0} \rightarrow n\pi$ the cross section $\sigma_\text{s-wave} \rightarrow 0$, which is an anti-resonance and corresponds to the valleys in Fig.~\ref{fig:2Dab} and Fig.~\ref{fig:3Dab}. The reduced cross section at anti-resonances in the parameter space is responsible for evading some experimental limits on the DM-baryon scattering cross section. The anti-resonances are not associated with any bound state. For an extended potential we have similar resonances and anti-resonances, they just appear at different $b$ values. In general, the location 
of the (anti-)resonance is a function of $c/b$.  

\begin{figure}
\centering 
\includegraphics[width=.8\textwidth]{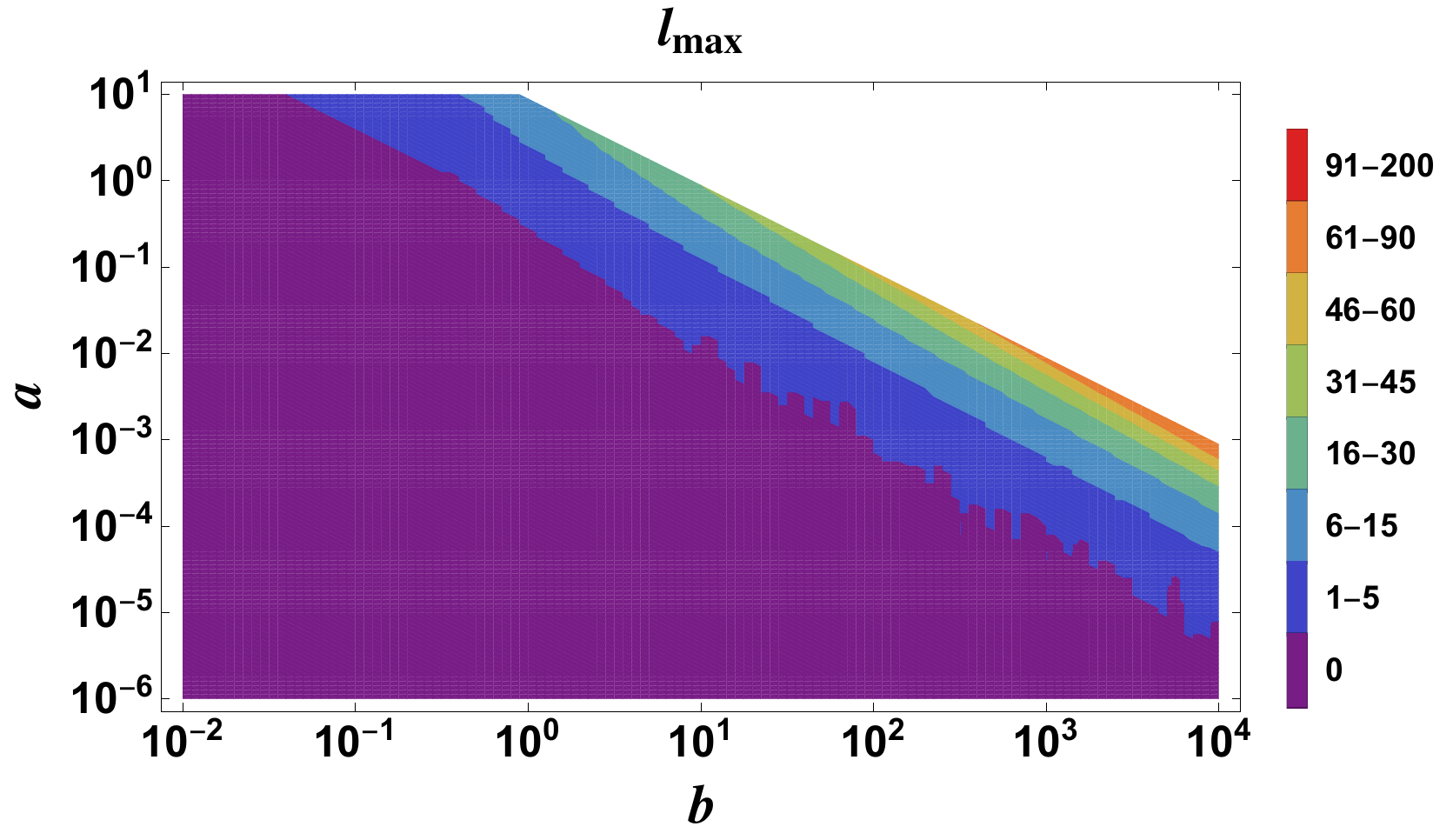}
\caption{\label{fig:Lmax} $l_\text{max}$ in the $(a,b)$ plane}
\end{figure}

\paragraph{S-wave Dominance:} It is usually the case that at low energy, s-wave scattering is dominant. In terms of $(a,b)$ this means small $a$. Figure~\ref{fig:Lmax} shows $l_\text{max}$, such that the contribution to the total cross section of partial waves from $l=0$ to $l=l_\text{max}$ is more than 99\%. We see that for a pure Yukawa interaction with $ab \ll 1$, the scattering is always s-wave dominated and quantum mechanical.  For an extended potential, s-wave dominance also requires $ac \ll 1$.  This is however also the condition for quantum resonant scattering, so the cross section for the extended potential in the resonant regime is automatically s-wave dominated and hence isotropic in the center of mass frame. This simplifies the expressions for the event rate of DM direct detection experiments.

\begin{figure}
\centering 
\includegraphics[width=.8\textwidth]{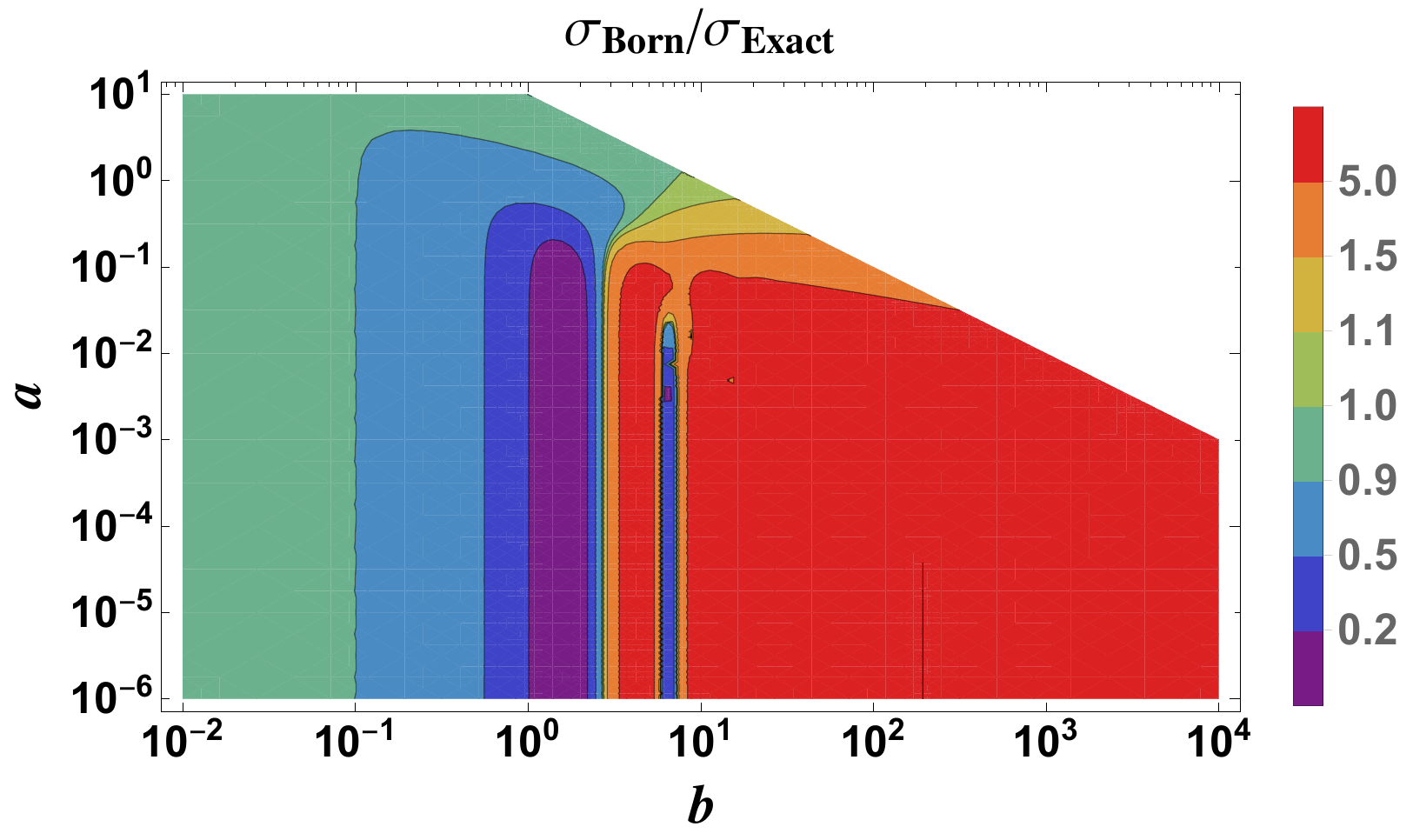}
\caption{\label{fig:BornExact} Ratio of $\sigma_\text{Born}$ over $\sigma_\text{Exact}$}
\end{figure}

\paragraph{Born Approximation Validity:}
Born approximation applies when $b \ll 1$. To quantify this, Fig.~\ref{fig:BornExact} shows the ratio of cross sections calculated by Born approximation and by numerical solution. Born approximation is within $\pm 10\%$ of the exact result for $b \lesssim 0.1$. This is also generally true for the extended potential. 

\begin{figure}
\centering 
\includegraphics[width=.8\textwidth]{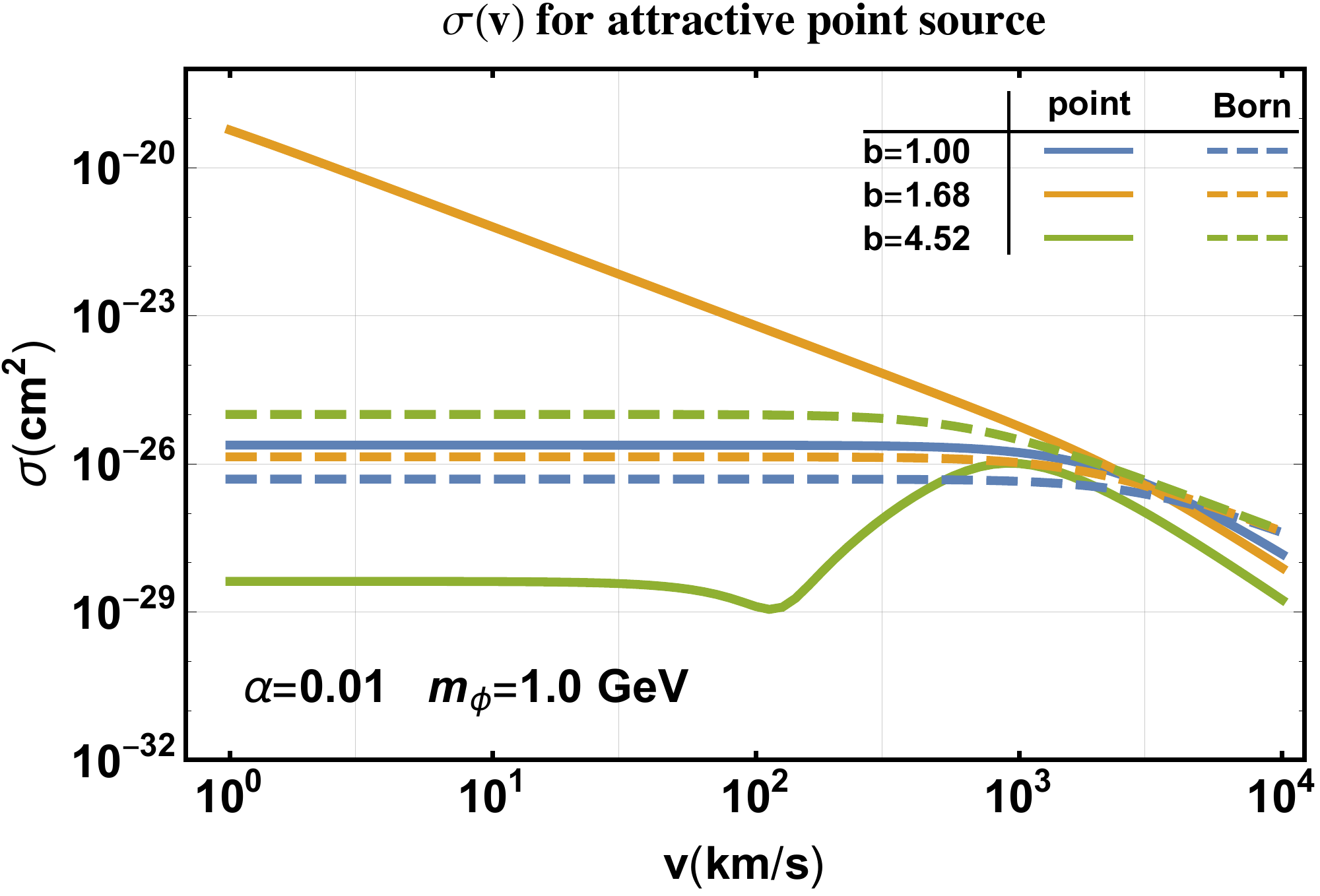}
\caption{\label{fig:Vdep} Velocity dependence of the cross section. $b=1.68$(tan) is on resonance and $b=4.52$(green) is on anti-resonance.}
\end{figure}

\paragraph{Velocity Dependence:} Another feature in the quantum resonant regime is that the cross section may have non-trivial velocity dependence, whereas Born approximation generally has no velocity dependence at small velocity. 
Figure~\ref{fig:Vdep} shows $\sigma m_\phi (v)$ for some illustrative values of $b$. At small velocity we have several different behaviors:\\
$\bullet$ On resonance ($b=1.68$), $\sigma \sim v^{-2}$, i.e., greatly enhanced at small $v$.  Here, s-wave unitarity fixes the cross-section at the peak of the resonance.  Using \eqref{eq:cxtotSwave}, with sin$^2(\delta_0) = 1$ gives
\begin{equation}
\label{eq:sigma_res}
    \sigma_{\rm res} = \frac{4 \pi}{\mu^2 \, v^2} = 4.9\times 10^{-21}\,{\rm cm}^2 \, \left(\frac{\rm GeV}{\mu}\frac{10^{-3} c}{v}\right)^2
\end{equation}
with negligible contribution from other partial waves.  Note the cross section becomes independent of $A$, coupling, and source size.\\
$\bullet$ 
At anti-resonance ($b=4.52$), the s-wave contribution to $\sigma$ vanishes and only the very small higher partial waves contribute.  The anti-resonances are in general away from any resonance for a different $A$ or $l$ and are therefore independent of velocity. \\
$\bullet$ 
For $b$ values well-separated from resonance and anti-resonance, $\sigma \sim v^0$ (independent of velocity) up to a large velocity whose value depends on parameters.  At such large velocities there is no simple expression for a general source distribution, but for a point Yukawa the scattering becomes Coulomb-like, so at sufficiently high $v$, $\sigma \sim v^{-4}$ for all $b$.  

Near but not on a resonance, there is a transition in behavior around $v\sim v*$ with $\sigma \sim v^{-2}$ for $v \gg v^*$ and $\sigma \sim $ constant for $v \ll v^*$.  In Appendix ~\ref{app:vdep} a useful approximate expression for the velocity dependence near resonant values of $\alpha$ is derived, valid when the radius $r_0$ of the extended Yukawa is large enough that it can be approximated as a square well.  Near the first resonance:

\begin{equation}
\label{eq:hi_v}
			\sigma \rightarrow 4 \pi / (\mu v)^2 ~~~~~~~~~~~~~~~~ v \gg v^* 
\end{equation}
and
\begin{equation}
\label{eq:lo_v}
			\sigma \rightarrow  \frac{8 m_\phi^2 r_0^3}{3 \pi \mu  \left(\sqrt{\alpha} - \sqrt{\alpha_{\rm res}}\right)^2 }~~~~ v \ll v^* ~
\end{equation}
with
\begin{equation}
\label{eq:v*}
v^* \equiv \frac{\pi |\sqrt{\alpha} - \sqrt{\alpha_{\rm res}}|}{m_\phi r_0 \sqrt{2\mu r_0/3}}~,
\end{equation}
and
\begin{equation}
 \alpha_{\rm res}  \equiv \left( \frac{\pi}{2} \right)^2 \frac{m_\phi^2 r_0}{6 \mu}.
\end{equation}

\subsection{Application to Experimental Constraints}
In the general spin-independent problem, the event rate per unit recoil energy is proportional to the differential cross section and one is interested in
\begin{equation}
\label{eq:rate}
\frac{dR}{dE_r} \sim \frac{d \sigma_A}{d \Omega} ~. 
\end{equation}
Here $d \sigma_A / d \Omega$ is the spin-independent DM-nucleus differential cross section for atomic mass $A$,  $E_r= q^2/2 m_{A}$ is the recoil energy and $q = 2\mu \, v\, \sin{\frac{\theta}{2}}$ is the momentum transfer.  Up until now, Born approximation has been almost universally assumed for analyzing experimental limits\footnote{An exception is \cite{Bai:2009cd}, which considered the possibility that DM resonant scattering could reconcile the DAMA and CDMS limits.}.  If Born approximation is valid, there are three distinct simplifications:\\

\paragraph{Form Factor and Size of Nucleus:} The difference between an extended source and a point-like source can be encoded in a form factor, $F(q)$, defined via
\begin{equation}
\label{eq:ff}
\frac{d \sigma}{d \Omega}_{ext} =  \frac{d \sigma}{d \Omega}_{pt} \, F^2(q)~,
\end{equation}
where in Born approximation it can be shown that $F(q=0) = 1$.  In a typical DM detection experiment with $v = \mathcal{O}(10^{-3}\,c)$ and nuclear size $\lesssim$ few fm, $q_{\rm max}\, r_0 \ll 1$.  Therefore, if Born approximation is valid, $F(0) = 1$ means that the cross section for the extended nucleus is the same as for a point-like nucleus.  (If $q$ is larger and higher partial waves beyond s-wave need to be taken into account, the $q$ dependence of the form factor can be calculated for any given model of the radial wave function of the nucleus, e.g., the Helm form factor~\cite{Helm:1956} is a popular choice.) 

\begin{figure}
\centering 
\includegraphics[width=.8\textwidth]{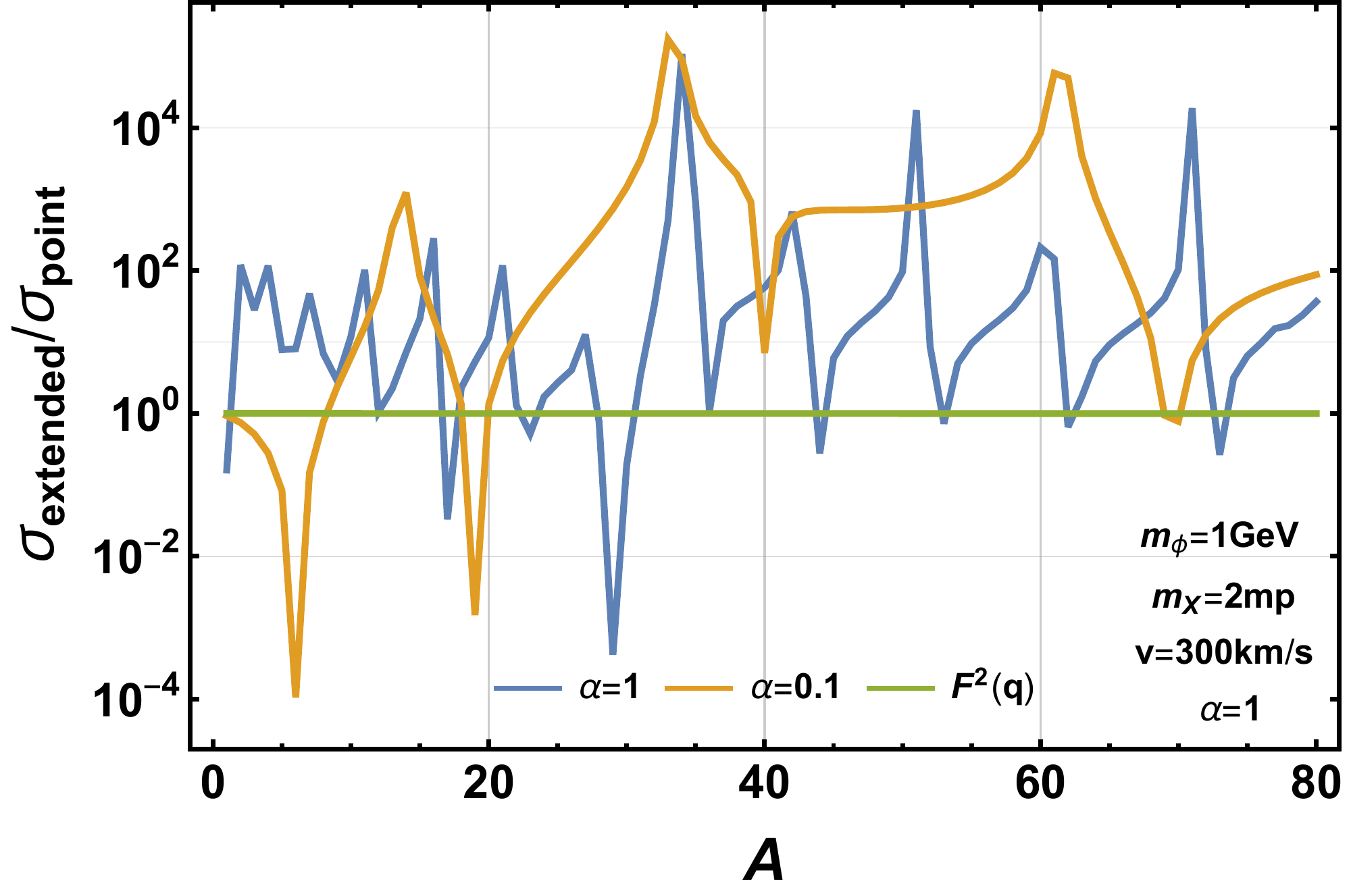}
\caption{\label{fig:formfactor} $\sigma_A^\text{extended}/\sigma_A^\text{point}$, as a function of $A$, comparing to the Helm form factor $F_A^2(q)$ used, e.g., in~\cite{Mahdawi:2017cxz}, for the mean value of $q$ given the scattering parameters.  The Helm form factor is essentially 1 and incapable of accounting for the overall scaling of the cross section coming from the finite size of the nucleus.   Additionally, for a heavier DM particle, $q$ can be large and the corresponding wavelength can be smaller than the size of the nucleus, even in the low-energy regime, yet $F_A(q)$ is still inaccurate compared to the numerical calculation.}
\end{figure}

The condition $F(0) = 1$ is however not general and can only be demonstrated in Born approximation.  Figure~\ref{fig:formfactor} compares the Helm~\cite{Helm:1956} form factor (green) to the ratio of the extended and point-like cross sections for two choices of $\alpha$, showing how dramatically the Born approximation, $F(0) = 1$, can fail in the quantum resonant regime -- here, by up to 5 orders of magnitude.

Finally, while we are most interested for the present work in the small $q$ limit where scattering is isotropic and $F(q)=\text{constant}$, we note that for the large $q$'s which can be encountered for massive DM and massive target nuclei, the angular dependence of the scattering cross section embodied in the Helm form factor will in general be  different from the true behavior, which must be determined by numerical calculation including higher partial waves. For large $A$, such that $A^{\frac{1}{3}} R_0$ is sufficiently large relative to $m_\phi^{-1}$ that a square-well approximation is adequate, analytic expressions again become possible.

\begin{figure}
\centering 
\includegraphics[width=.8\textwidth]{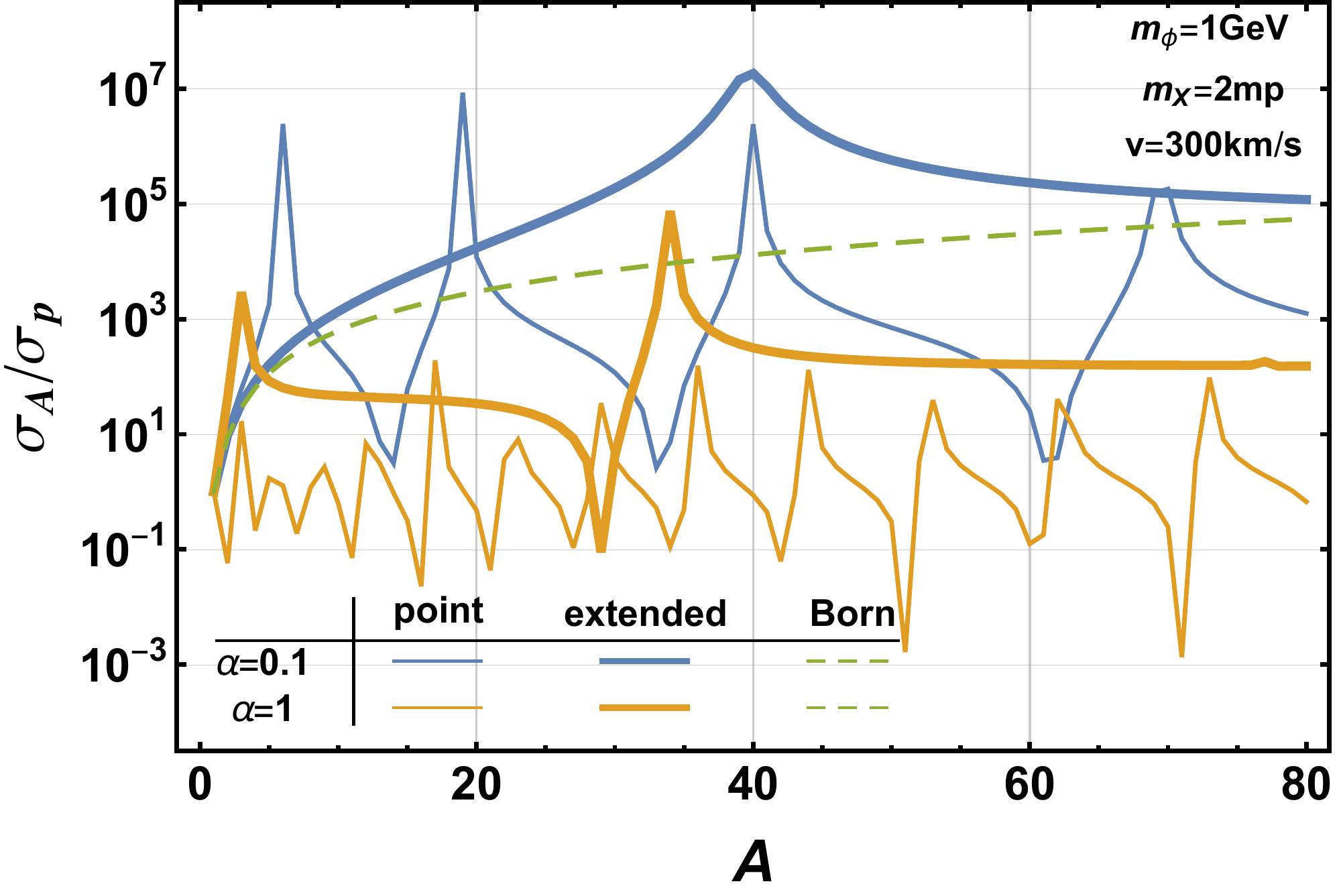}
\caption{\label{fig:Ascaling} $\sigma_A / \sigma_p$ as a function of $A$.  The green line is the Born approximation prediction~\eqref{eq:Ascaling}, while blue and tan lines are the result of numerical calculation for point source and extended source, respectively. For an extended potential we take the nucleus radius as $ = A^{\frac{1}{3}} \text{\,fm}$;  in all cases the coupling is $\alpha_A = A\alpha$. 
} 
\end{figure}

\paragraph{Scaling of $\sigma_A$ with $A$:} The commonly-assumed Born approximation relationship between the DM-nucleus cross section in Born approximation, $\sigma_A^{\rm Born}$, and the DM-nucleon cross section, $\sigma_p$, is: 
\begin{equation}
\label{eq:Ascaling}
\sigma_{A}^{\rm Born}=\sigma_{p}\left(\frac{\mu_{A}}{\mu_{p}}\right)^{2} A^{2}.
\end{equation}
This can be obtained from Eq.~\eqref{eq:sm-cxtotborn} with $ab \ll 1$ or $\mu v \ll m_\phi$, i.e. in the low energy regime compared to $m_\phi$. When eq.~\eqref{eq:Ascaling} is valid, as is the case for WIMP experiments~\cite{Lewin:1996}, the final result of an experiment can be reported as a limit for $\sigma_p$.  It is for this reason that different experiments or observations can put universal limits on $\sigma_p$ and compare with each other despite the fact that they are using different target nuclei.   
However the scaling relationship~\eqref{eq:Ascaling} between $\sigma_A^{\rm Born}$ and $\sigma_p$ does not work in the resonant regime, as is shown in Fig.~\ref{fig:Ascaling}. In fact, the ratio $\sigma_A / \sigma_p$ becomes highly parameter dependent. As a consequence, there is no universal rule to convert an experimental limit on $\sigma_A$ to a single parameter $\sigma_p$, and it is non-trivial to compare the results of different experiments.   An additional complication is that experiments involving multiple materials with different $A$ require an even more subtle analysis.\\ 

\begin{figure}
\centering 
\includegraphics[width=.8\textwidth]{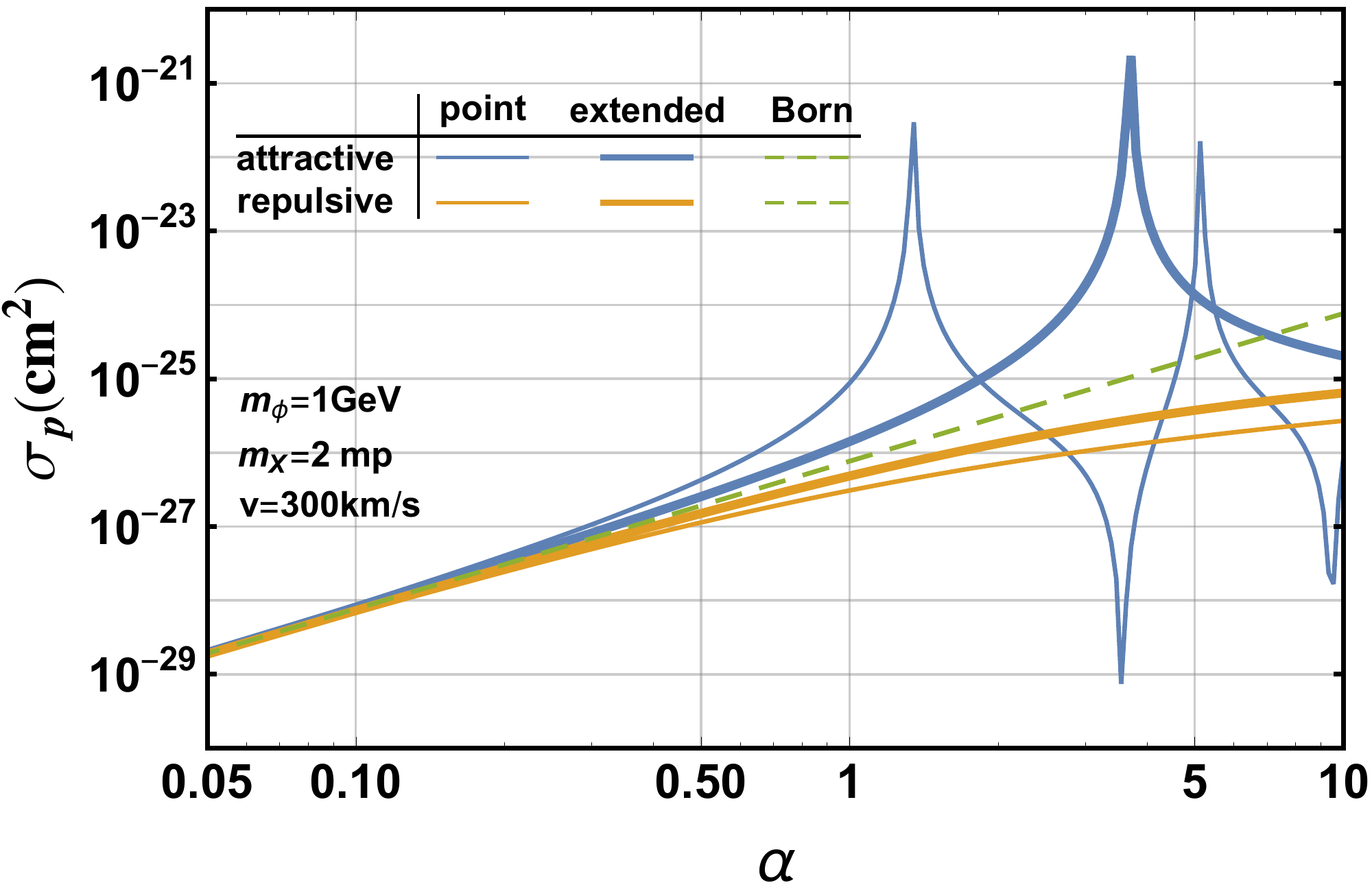}
\caption{\label{fig:ptextborn} $\sigma_p$ as a function of $\alpha$ for: Born approximation, point source and extended source. Born approximation only holds for $\alpha \lesssim 0.3$, where the size of the nucleus and the sign of the potential do not matter.}
\end{figure}

\paragraph{Connection to $\sigma_p$:} In the absence of the Born approximation scaling relationship embodied in eq.~\eqref{eq:Ascaling}, the only way to relate $\sigma_A$ to $\sigma_p$ is to solve for both, under a given assumption for ($\alpha,\, m_X,\, m_\phi$).  Neither $\sigma_A$ nor $\sigma_p$ are in general calculable perturbatively, and even for point-like proton, $\sigma_p$ is not given by the simple Born approximation expression for $\sigma_p$.  In Fig.~\ref{fig:ptextborn} we compare both point-like and extended solutions of the Schrödinger equation to Born approximation for DM-proton scattering.  Our extended model of the proton takes it to be a sphere of radius 1 fm (rms charge radius 0.77 fm) sourcing the Yukawa potential.

We have shown in Figs.~\ref{fig:formfactor}-\ref{fig:ptextborn} that Born approximation fails badly in the quantum resonant regime, in all three respects -- sensitivity to size, $A$ dependence, and dependence of $\sigma_p$ on fundamental parameters -- so we must change the way the experiments are interpreted.  We discard the form factor in our analysis and model the finite size of the nucleus with full numerical calculation for the extended Yukawa potential.  We have already shown that in the quantum resonant regime we are considering, the scattering is s-wave dominated and isotropic.

\begin{figure}
\centering 
\includegraphics[width=.7\textwidth]{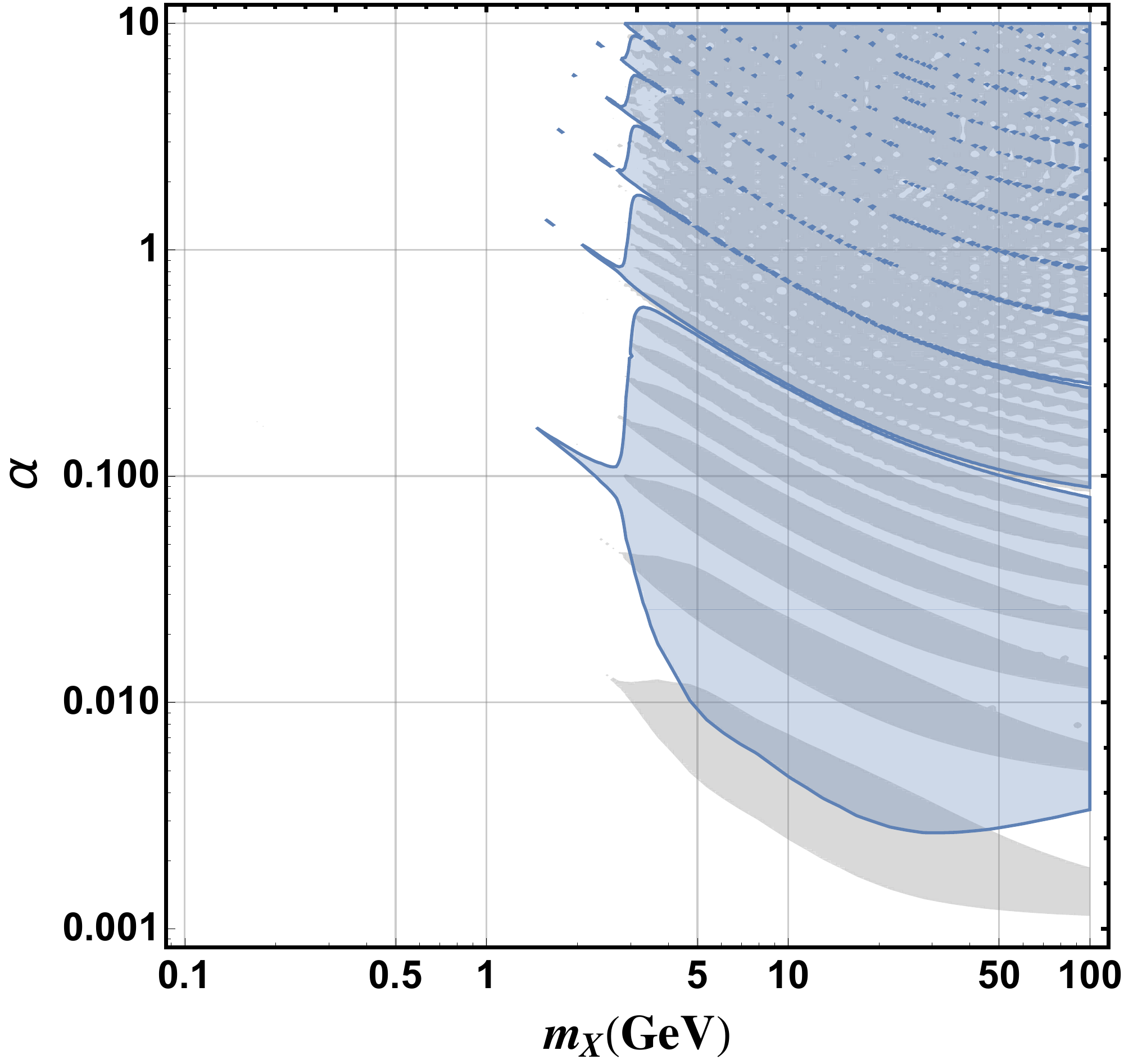}
\caption{\label{fig:XQCalphamx} Exclusion region in the ($\alpha$,$m_X$) plane from XQC, for attractive point source (gray) and attractive extended source (blue) taking $R_0 =1.0 \text{\,fm}$ and $m_\phi=1\text{\,GeV}$.}
\end{figure}

\section{\label{sec:reinterpret}Reinterpreting Direct Detection Experiments} 
In this and the following section we derive the limits from various experiments and observations, focusing on the case of an attractive DM-nucleon interaction except where noted, for which the analysis is generally more subtle.
\subsection{\label{sec:reinterpretXQC}The XQC Experiment}The X-ray Quantum Calorimeter (XQC)~\cite{McCammon:2002} was an experiment intended to measure the diffuse x-ray background using micro-calorimeters on board a rocket sent to about 100 km altitude in the atmosphere. The results can also be used to put limits on the DM-nucleon cross section and extensive studies have been performed~\cite{Wandelt:2000ad,Zaharijas:2004jv,Erickcek:2007jv,Mahdawi:2017cxz,Mahdawi:2018euy}. However these analyses uniformly used the non-valid Born approximation to extract limits on $\sigma_p$, so in this section we re-interpret the latest analyses of XQC limits~\cite{Mahdawi:2017cxz,Mahdawi:2018euy} as required to obtain reliable limits in the quantum resonant regime.  The procedure is:
\begin{itemize}
   \item Rescale the limits reported for $\sigma_p$, back to the limits on the actual cross section that XQC is constraining -- $\sigma_{28}$ -- undoing the assumed Born approximation scaling, ~\eqref{eq:Ascaling}. The XQC detector is actually made of silicon and a thin HgTe film, but the latter makes an insignificant contribution and we ignore it for simplicity. For DM mass $\sim$ few GeV, of special interest in connection with sexaquarks, the maximum momentum transfer is very small and the form factor $F(q)$ applied in~\cite{Mahdawi:2017cxz,Mahdawi:2018euy} is essentially 1.  The dashed curve XQC($\sigma_{28}$) in Fig.~\ref{fig:experiments} shows the resultant limits on $\sigma_{28}$. 
   
      \item Using the numerical solution to the Yukawa potential model for 
      extended nuclei, 
      calculate $\sigma_{28}$ everywhere in the parameter space $(\alpha, m_X, m_\phi) $. 
      We adopt $=R_0\,A^{\frac{1}{3} }$, and calculate for $R_0 = 1.0$ and $1.2\text{\,fm}$ to assess the sensitivity to the exact size of the nucleus. Comparing to the observational limit on $\sigma_{28}$, we then obtain the excluded region in $(\alpha, m_X, m_\phi) $ shown in Fig.~\ref{fig:XQCalphamx}. Due to the resonant behavior, the excluded/allowed regions of the parameter space has islands and holes whose exact positions depend on $R_0$.  
      
      \item Calculate $\sigma_{p}$ for the allowed values of $(\alpha, m_X) $ to find the allowed region in the $(\sigma_{p}, m_X)$ plane, for a given choice of $ m_\phi$.  This enables a standardized comparison to other limits.   
      
\end{itemize}

\begin{figure}
\centering 
\includegraphics[width=1.0\textwidth]{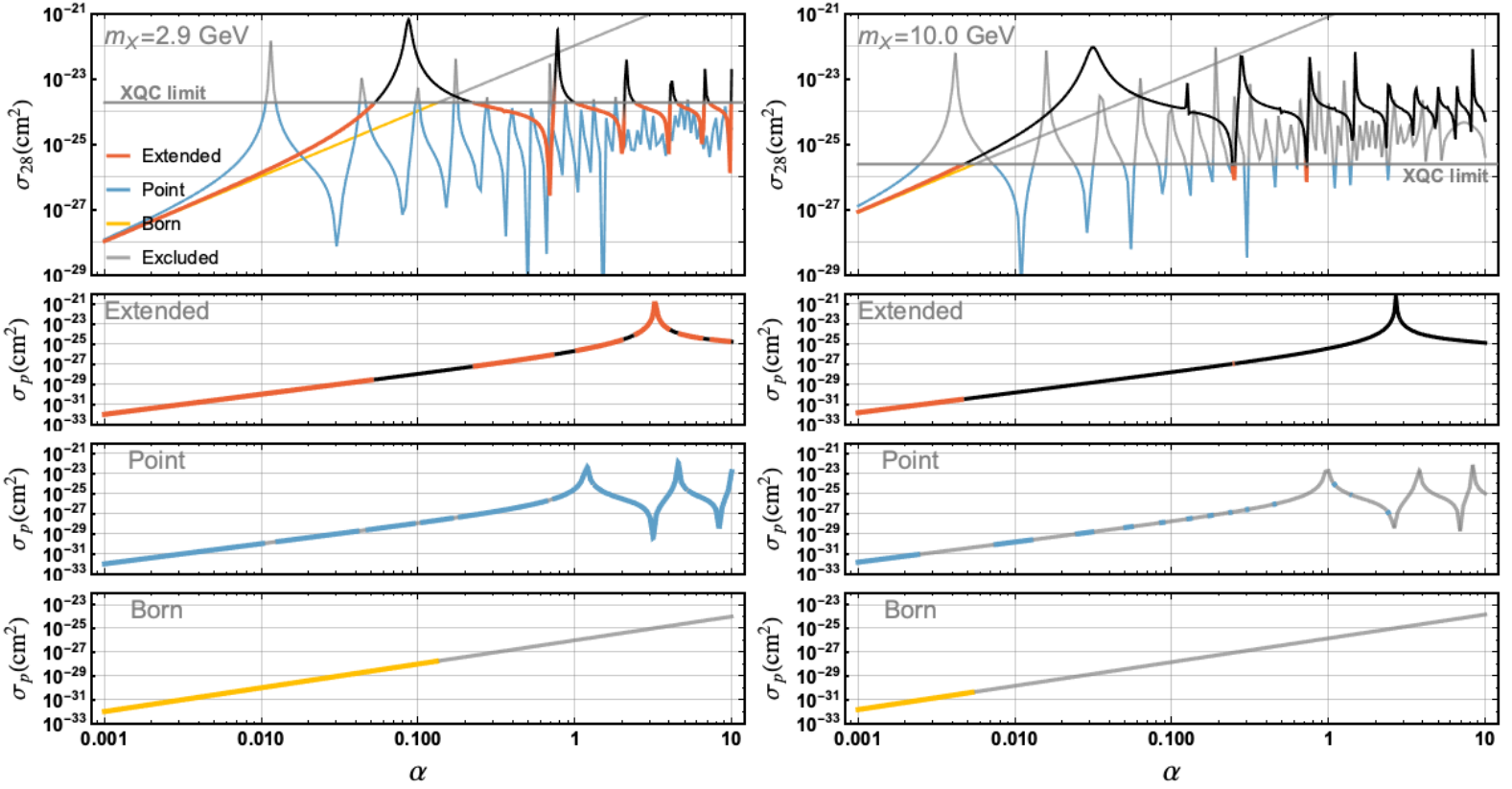}
\caption{\label{fig:mxcombined}  The top panel shows $\sigma_{28}$ versus $\alpha$ in the attractive case, for $m_X = 2.9$ GeV (left) and $10$ GeV (right).  The red/black lines show the exact treatment with extended nucleus, while cyan/grey and orange/grey show the results for point nucleus and Born approximation.  The grey horizontal line is the XQC limit on $\sigma_{28}$, so regions of $\alpha$ in which the predicted $\sigma_{28}$ exceeds this limit are excluded and the curves are correspondingly shown in black or grey. The lower panels show, in the same color scheme, the corresponding $\sigma_{p}$ predictions.  The black regions are excluded; the grey regions would be excluded if the point Yukawa or Born approximations were applicable. This shows how an XQC upper limit on $\sigma_{28}$ maps into excluded regions of $\sigma_p$, and illustrates how there can be "islands" of allowed and excluded parameters, unlike in Born approximation or to a limited extent for the point Yukawa shown in the lower two rows.}
\end{figure}

The procedure and its non-trivial character are illustrated in Fig.~\ref{fig:mxcombined}, which shows the exact and Born predictions for $\sigma_{28}$ and $\sigma_p$ as a function of $\alpha$, for $m_X= 2.9$ and 10 GeV. Focusing first on the top panel, one sees how the non-perturbative cross-section exceeds the Born approximation in some regions of $\alpha$ but is below it in others, and how important it is to take into account the extended nucleus properly.  We return in Sec.~\ref{sec:results} below to how these changes impact the XQC exclusion region in $\sigma_p$.

As noted in Sec.~\ref{sec:GenResNonPert} and illustrated in Fig.~\ref{fig:Vdep}, the full, non-perturbative cross section has a non-trivial velocity dependence near a resonance, behaving as $v^{-2}$ until saturating at some minimum velocity determined by the distance from resonance. We treat this non-trivial velocity dependence as follows.  For XQC, we obtain a preliminary exclusion region evaluating $\sigma_{28}$ at the characteristic DM velocity of $v=300 \text{\,km/s}$ and then check near the resonances whether a more accurate treatment is required.
The maximum (escape) velocity for DM particles in the Milky Way halo is roughly $v \sim 600 \text{\,km/s}$. Thus near a resonance, the $v^{-2}$ behavior of the cross section can result in a cross section as small as one-fourth of the value at $300 \text{\,km/s}$. 
For $m_X \gtrsim 3 \text{\,GeV}$, this smaller cross section is still excluded by XQC because the basic XQC limit on the number of DM scattering events is so stringent.  (For details, see~\cite{Mahdawi:2018euy}.) This can be seen directly from the right panel of Fig.~\ref{fig:mxcombined} for $m_X=10\text{\,GeV}$, where the XQC limit is usually more than three orders of magnitude smaller than the predicted cross section near the resonance.   At still larger mass, DM can trigger the detector at velocities below $300 \text{\,km/s}$ with a equal or larger cross section than that at $300 \text{\,km/s}$, so the spectrum-weighted event rate can potentially actually be higher than calculated using a constant $v=300 \text{\,km/s}$ value. In that case, the limits presented are conservative, as desired.

For smaller DM mass, $m_X \lesssim 3 \text{\,GeV}$, as shown in the left panel of Fig.~\ref{fig:mxcombined}, the XQC limit on $\sigma_{28}$ is closer to the calculated cross section near or on resonance, and a factor-4 smaller cross section at $v \sim 600 \text{\,km/s}$ could potentially evade the XQC limit. This would shrink the small blue peninsulas in Figs.~\ref{fig:XQCalphamx}, ~\ref{fig:alphamxAll} (left), and~\ref{fig:sigmxAll} (left) for $m_X \lesssim 3\text{\,GeV}$.  However the peninsula is an uncertain region anyway due to its sensitivity to details of the nuclear wave function, which is not perfectly well-determined.  Therefore we do not attempt a more refined analysis and simply leave the peninsula region unfilled in the cross-section plots, to indicate the limit is uncertain. 


\subsection{The CRESST Surface Run Experiment}
\begin{figure}
\centering 
\includegraphics[width=1.0\textwidth]{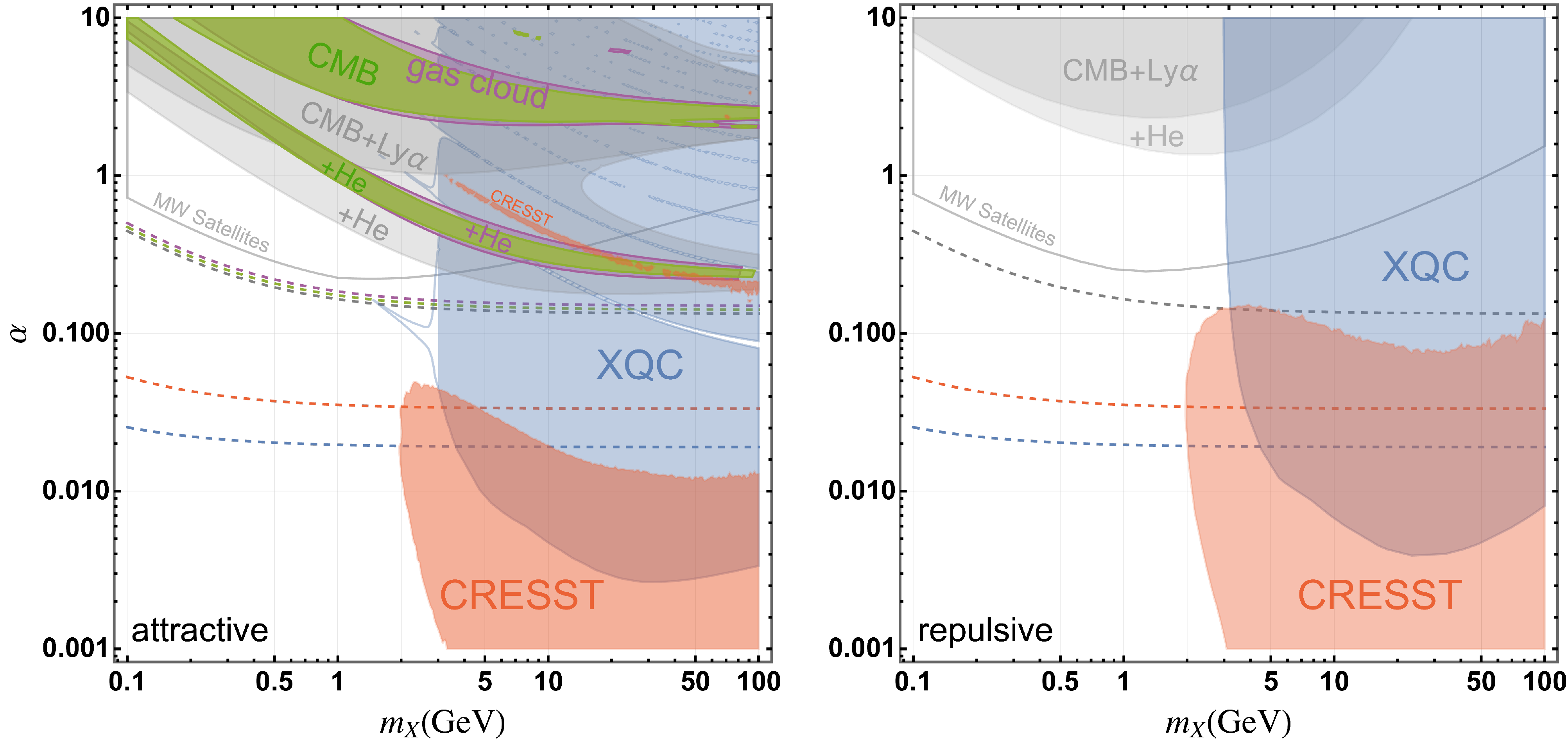}

\caption{\label{fig:alphamxAll} Exclusion region in the ($\alpha$,$m_X$) plane for $m_\phi=1 \text{\,GeV}$, from XQC (blue) and CRESST (red), for attractive (left panel) and repulsive (right panel) interactions;  the boundaries should be considered blurry because of their sensitivity to the details of the nuclear wave function. Note the narrow island of large $\alpha$ excluded by CRESST for $m_X\gtrsim 3$ GeV in the attractive case, thanks to anti-resonance behavior in this regime.  The secure limit from the CMB in the attractive case is shown in green, and the nearly-identical limit from gas clouds in purple. In each pair, the upper wider bands show the limit including only H, while the lower narrower stripes labeled ``+He" are excluded due to the contribution of He.  There are no corresponding CMB and gas cloud  limits in the repulsive case, because there is no value of $\alpha$ giving a cross section large enough to saturate the bounds. If the limits based on combining CMB with Ly-$\alpha$ are trustworthy (see \cite{Hui:2016ltb,vandenBosch+Shattering19} for cautions), then the gray regions can be excluded using the results from \cite{Xu:2019}, where the upper darker region includes H only and the lower lighter gray region is excluded by He in our non-perturbative analysis.  In the left panel for an attractive interaction, the unconstrained gap region above $m_X\approx$ 15 GeV is due to an anti-resonance in DM-He scattering.  The gray solid line indicates the upper limits of $\alpha$ if the recent constraints based on Milky Way satellites~\cite{DES:2020fxi} are validated. For each experiment or observation, the dashed line in the same color indicates the parameter values such that the dimensionless parameter $b$ equals one: for $A=4 $ relevant for He in CMB, gas cloud, and Ly-$\alpha$; $A=16 $ for $O$ in CRESST; $A=28 $ (XQC). When $b \lesssim 1$ Born approximation is reliable for the point-like and extended Yukawa model, however unless $b \ll 1$ the extent of the nucleus still matters. If the interaction is repulsive (right panel), the boundary of XQC exclusion region is smooth due to the lack of resonance and there is no anti-resonance contribution to the CRESST limit. }

\end{figure}

The CRESST 2017 surface run took data using a cryogenic detector operated
by the CRESST collaboration~\cite{Angloher:2017sxg,Strauss:2017cuu,Strauss:2017cam} near ground level. The detector is made of  $\mathrm{Al}_{2} \mathrm{O}_{3}$ and observed a total of 511 nuclear recoil events. It is shielded by $\sim$ 30 cm of concrete in addition to the atmosphere of Earth. As a result, DM with large cross section in the overburden will loose too much energy to register in the detector.  Roughly speaking, this places a \textit{maximum} value on the cross section $\sigma_p$ which can be probed in the CRESST surface run.  Since the detector, the concrete and the atmosphere contain nuclei with different $A$, we modified the analysis done in~\cite{Mahdawi:2018euy}, calculating each of the cross sections required in the analysis for any given set of parameters ($\alpha,m_\phi,m_X$).  We thereby obtain the limits on the ($\alpha,m_X$) parameter space shown in Fig.~\ref{fig:alphamxAll}. The non-trivial character of the new exclusion limits is already evident in the emergence of a (narrow) excluded island which appears at large $\alpha$ for attractive interaction, from about $\alpha \approx1$ at $m_X \approx 3\text{\,GeV}$ to $\alpha \approx 0.2$ at $m_X =100 \text{\,GeV}$. This excluded band arises because even though the fundamental coupling is large, in this  parameter regime the relevant nuclear scattering cross-sections are small due to the existence of an anti-resonance.  In this region, DM can reach the CRESST detector with sufficient energy to be detected, while for larger and smaller parameter values it loses too much energy in the overburden. 

For CRESST, the lower exclusion region is close to the Born regime while the upper island for the attractive interaction is on anti-resonance for most elements involved. Both regions are far from any resonance so that our treatment ignoring the velocity dependence near resonance is accurate. 

\section{\label{sec:CMBast} CMB and Astrophysical Constraints}
Another class of constraints, which extend to lower DM mass but are less powerful in terms of cross-section limits than the direct detection experiments discussed in the previous section, derive from limits on heat exchange and friction between baryons and DM.  The pioneering work of \cite{Dvorkin:2013cea} showed that these effects from DM-baryon interactions suppress the cosmological growth of density contrast on small scales, so that precision data on the CMB power-spectrum constrains the DM-baryon cross section. 
A related but independent constraint comes from limits on heating of cold, dense and long-lived gas clouds in the Milky Way \cite{Wadekar:2019mpc}.  

The CMB and gas clouds constraints have two features in common: {\it i)}  H and He are present in the cosmological abundance ratio and {\it ii)} the cross section values which can be constrained are in the non-perturbative regime, where Born scaling with $A$ is not valid and finite size effects are important.  As we shall see, the correct treatment significantly changes the derived limits. Two other systems have been proposed to probe structure growth on still smaller scales than accessed by the CMB power spectrum:  Ly-$\alpha$ forest and dwarf satellites in the Milky Way.  Using the reported Born approximation limits from \cite{Xu:2018efh,DES:2020fxi}, we provide the corresponding exact limits; if the constraints of \cite{Xu:2018efh,DES:2020fxi} are established as robust, these will be stronger than the CMB and gas clouds limits\footnote{In a preliminary posting of this paper (arXiv:2101.00142v1\cite{Xu:2020qjk_v1}) we only analyzed the CMB+Ly-$\alpha$ limits and did not consider the more robust but difficult-to-treat pure CMB limit.  Here they are both given and shown separately in the figures, now also treating the repulsive case.}. 

The observational constraints considered in this section bound a linear combination of $\sigma_{\rm H}$ and $\sigma_{\rm He}$, weighted by the H and He abundances and the energy- or momentum-transfer efficiency in a DM-nucleus collision.  In the case of Milky Way gas clouds, higher mass nuclei contribute as well.  Analyzing using Born approximation, as was done in previous analyses, implies assuming the fixed cross section ratio  $\sigma_{A}^{\rm Born}=\sigma_{p}\left(\frac{\mu_{A}}{\mu_{p}}\right)^{2} A^{2}$.  For instance  for $m_X = \{2,\,10\}$ GeV, $\sigma_{H e}=16\, \frac{\mu_{He}^{2}}{\mu_{H}^{2}} \, \sigma_{H} = \{66,\,197\}\, \sigma_{H} $ (Eq.~\eqref{eq:Ascaling}).  Thus in Born approximation He plays an important role, even though the He abundance is only $\approx$ 10\% that of H,  due to the large ratio of Born cross sections.  For high DM mass, the importance of heavier nuclei relative to protons is further enhanced by the higher energy transfer efficiency in collisions of more nearly equal masses, with the average fraction of energy transferred to a slow-moving nucleus being $2 m_A m_X/(m_A + m_X)^2$.  

However in reality, for much of parameter space in the non-perturbative regime, $\sigma_{He}$ is actually much smaller than the Born approximation estimate.  The essential point is that to get cross sections at the barn level requires both strong coupling ($\alpha \approx \mathcal{O}(1)$) and being near a resonance value of $\alpha$.  But the resonances of H and He are at quite different values of $\alpha$, so the H and He cross sections do not scale together as they do in Born approximation.  This is demonstrated in detail below.

\begin{figure}
\centering 
\includegraphics[width=.8\textwidth]{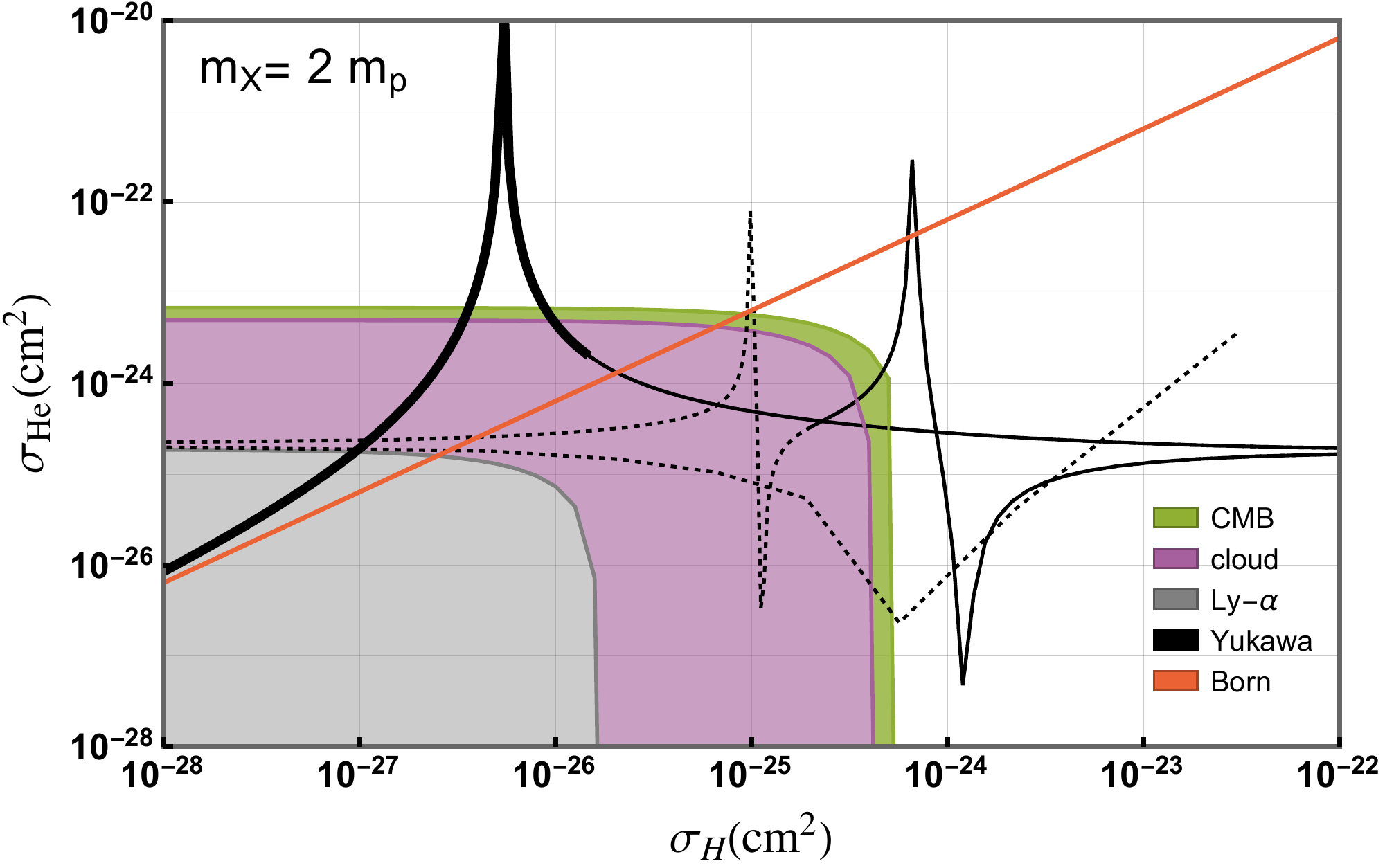}
\caption{\label{fig:sigsig} The black line is the parametric curve of $\{\sigma_H,\sigma _{He}\}$ for an attractive interaction and  $m_X=2m_p$, as a function of $\alpha$, starting at the lower left (heavy line) $\alpha$ increases from 0.1 to 1; the medium line is for $1<\alpha <10 $ and the dashed line for $10< \alpha < 30$. The red line is the Born approximation relationship, $\sigma_{H e}=16 \frac{\mu_{He}^{2}}{\mu_{H}^{2}} \sigma_{H}$, and the solid regions show the allowed domain from the CMB, gas clouds and CMB-Ly$\alpha$.}
\end{figure}
    
\subsection{\label{sec:CMB}Constraints from the CMB}
DM interacting with baryons in the early universe leaves an imprint on CMB 
observations, allowing limits on the cross section to be obtained~\cite{Dvorkin:2013cea, Gluscevic:2017ywp, Xu:2018efh}. We investigate the most recent results from~\cite{Xu:2018efh} (Fig.~\ref{fig:experiments}) where both DM-Hydrogen and DM-Helium scattering are included. However~\cite{Xu:2018efh} uses Born approximation and assumes that only the proton interacts with DM in the initial calculation. It is clear now that any probes of the interaction of DM and baryons, with $A > 1$ in the non-perturbative regime, suffer from the problem of breaking down of the Born scaling~\eqref{eq:Ascaling}.  So a reinterpretation as done above for the XQC and CRESST experiments is necessary. Since the Yukawa cross section is velocity independent except near a resonant value of $\alpha$ , we begin by examining the $n=0$ case treated in Sec. VII of ~\cite{Xu:2018efh}; in their notation $\sigma(v) = \sigma_0 (v/c)^{n=0}$.

In the early universe both DM-H ($A=1$) and DM-He ($A=4$) interactions are significant.  What governs the limits on the cross section is the momentum transfer rate between DM and baryon fluids.  Following the notation of ~\cite{Xu:2018efh} and starting from their Eq. (13),
\begin{align}
\label{RX}
R_X
&=ac_0\sum_{i}\frac{\rho_i \sigma_i v_i}{m_X+m_ i} \nonumber \\ 
&=\frac{ac_0\rho_b v_H}{m_X + m_H} \sigma_H(1-Y_{He})   \left( 1+ \frac{Y_{He}}{1-Y_{He}} \frac{\sigma_{He}}{\sigma_H} \frac{v_{He}}{v_H}\frac{m_X+m_H}{m_X+m_{He}}  \right)~~,
\end{align}
where $a \text{ and }c_0$ are constants not relevant for us, the sum on $i$ is the sum over all baryon species (here H and He), $m_i$ is the nuclear mass, $\rho_i$ is the density of nucleus $i$, and $v_i \text{ and } \sigma_i$ are, respectively, the average relative velocity and the scattering cross section between DM and baryon $i$; $Y_{\text{He}} = 0.24$ is the Helium mass fraction used by~\cite{Xu:2018efh}. Defining $\sigma_0$ as the conservative limit on $\sigma_H$ when $\sigma_{He}=0$, ref~\cite{Xu:2018efh} reports $\sigma_0$ as a function of $m_X$. To include DM-He scattering, noticing that $v_{He}/v_H = \sqrt{\mu_H/\mu_{He} }\geq \frac{1}{2}$, we can write $R_X$ as 
\begin{align}
\label{RXHe}
R_X
&=\frac{ac_0\rho_b v_H}{m_X + m_H} \sigma_0(1-Y_{He})\nonumber \\ 
&=\frac{ac_0\rho_b v_H}{m_X + m_H} \sigma_H(1-Y_{He})   \left( 1+ \frac{Y_{He}}{1-Y_{He}} \frac{\sigma_{He}}{\sigma_H} \frac{v_{He}}{v_H}\frac{m_X+m_H}{m_X+m_{He}}  \right) \nonumber \\ 
&\geq\frac{ac_0\rho_b v_H}{m_X + m_H} \sigma_H(1-Y_{He})   \left( 1+ \frac{Y_{He}}{1-Y_{He}} \frac{\sigma_{He}}{\sigma_H} \frac{m_X+m_H}{2(m_X+m_{He})}  \right)
\end{align}
which gives
\begin{equation}
\label{eq:sigHsig0}
\sigma_H + \left[ \frac{Y_{He}}{1-Y_{He}} \frac{m_X+m_H}{2(m_X+m_{He})} \right] \sigma_{He} \leq \sigma_0 ~. 
\end{equation}
Depending on the assumed $\sigma_{He} / \sigma_H$ one gets different limits on $\sigma_H$ from $\sigma_0$, when $\sigma_{He}$ is non-zero. The results of ref~\cite{Xu:2018efh} were obtained using
\begin{equation}
\label{bornXu}
\sigma_{H e}=4 \frac{\mu_{He}^{2}}{\mu_{H}^{2}} \sigma_{H}~~,
\end{equation}
corresponding to assuming DM only scatters on protons and not the neutrons in the Helium, which is not appropriate if DM is an isoscalar as is the case for sexaquark DM and other non-photon-mediated DM.  Eq.~\eqref{bornXu} also assumes the validity of Born approximation and thus breaks down in the quantum resonant regime. 

To place constraints on the Yukawa potential parameter space, we calculate ($\sigma_{He}$, $\sigma_H$) for a given choice of parameters ($\alpha,m_\phi,m_X$), then check whether the inequality eq.~\eqref{eq:sigHsig0} is violated~\footnote{Since the true DM-baryon cross section is velocity dependent near a resonance value of $\alpha$ for an attractive interaction, as discussed in Sec.~\ref{sec:GenResNonPert}, we evaluate ($\sigma_{He}$, $\sigma_H$) at the typical relative velocity in the most constraining epoch for the CMB and CMB-Ly-$\alpha$ limits: $v_{\rm{CMB}}\approx 40$ km/s and $v_{\rm{Ly}\alpha}\approx 110$ km/s, respectively.  These values can be inferred from the last two columns of Table I of \cite{Xu:2018efh}, by solving for the velocities such that the $n=0$ and $n=-2$ cross section limits are equal using the $m_X=1$ GeV limits in that table. The $v^{-2}$ velocity dependence only matters near the first resonance of $\sigma_{\rm{He}}$, which are the "+He" regions in Fig.~\ref{fig:alphamxAll} (left). However, even for the maximum relative velocity $\sim$ $2v_{\rm{CMB}}$ (or $2v_{\rm{Ly}\alpha}\approx 110\,{\rm km/s}$) is used, $\sigma_{\rm {He}}$ would only be 4 times smaller, which is still well excluded. This can be seen from Fig.~\ref{fig:sigsig}, where the $\sigma_{\rm{He}}$ peak is more than $10^3$ times above the allowed region. Thus our exclusion region is not modified by the velocity dependence, as a result of the strength of the constraints.}.  We again model the proton as a solid sphere of radius 1 fm, since its rms charge radius $\approx 0.8 \text{\,fm}$.   Figure~\ref{fig:alphamxAll} shows the resultant exclusion region in ($\alpha,m_\phi,m_X$) in green, based on the $m_X$-dependent limit of ~\cite{Xu:2018efh} from the CMB temperature and polarization power spectra; e.g., $\sigma_0<6.3 \times 10^{-25}\,{\rm cm^2}$ for $m_X = $ 2 GeV. For an attractive interaction, the narrow green band at lower $\alpha$ is where He is near resonance and the upper green exclusion region is where H is near resonance. There are also excluded "islands" at large $\alpha$ which are not shown. 

It is instructive to identify the origin of the various excluded and allowed $\alpha$ ranges seen in Fig.~\ref{fig:alphamxAll} (left).  Their origin can be understood by considering a parametric plot of $\sigma_{\rm H}-\sigma_{\rm He}$ as a function of $\alpha$, shown in Fig.~\ref{fig:sigsig} for the case of $m_X = 2 $ GeV.  The colored regions in Fig.~\ref{fig:sigsig} show the allowed regions from various constraints.  Following the thick black line from low $\alpha$, He passes through a resonance for $0.6 < \alpha < 0.8$, producing the lower excluded band in Fig.~\ref{fig:alphamxAll} (left).  For future reference, in this region $\sigma_H \approx 10^{-26.5}\, {\rm cm}^2$.  As $\alpha$ increases further, neither cross section is large enough to violate the bound until H is close enough to resonance for $\alpha \approx 3-5$,  that the gas clouds and CMB constraints are violated.  Fig.~\ref{fig:sigsig} also reveals an excluded region due to He for $\alpha \approx 15$.  Importantly, for $\alpha \approx 25$ (the region of the anti-resonance in H), $\sigma_H \approx 10^{-26.5}\, {\rm cm}^2$.  This means that the lower excluded band in Fig.~\ref{fig:alphamxAll} (left) around $ 0.6 < \alpha <  0.8$ for $m_X = 2$ GeV, \emph{does not} imply that the corresponding band around $\sigma_H \approx 10^{-26.5}\, {\rm cm}^2$ is excluded, because that same range of $\sigma_H$ can be produced by an allowed (albeit large) $\alpha$.  

\subsection{\label{sec:CMBplus}Potential constraints from Lyman-$\alpha$ forest and satellite dwarf galaxies}

 If the CMB+Lyman-$\alpha$ constraint of~\cite{Xu:2018efh} is valid (questioned in~\cite{Hui:2016ltb}), the allowed region in Fig.~\ref{fig:sigsig} (applicable for an attractive interaction) is reduced to the gray domain: $\sigma_0<1.7 \times 10^{-26}\,{\rm cm^2}$ for $m_X = 2$ GeV.  For small $\alpha$, $\sigma_{He}/\sigma_H$ is larger than given by Born approximation, so the correct non-perturbative limit on $\sigma_p$ is stronger than deduced using Born approximation. At larger mass, when $m_X \gtrsim 10-15\text{\,GeV}$, another non-perturbative effect gives rise to a gap appearing in the CMB+Lyman-$\alpha$ limit for relatively large $\alpha$ due to He anti-resonance regions in ($\alpha,m_\phi,m_X$), akin to those encountered in the XQC and CRESST analyses.

Interactions between DM and baryons which give rise to the cosmological CMB limits can also reveal themselves through suppression of low-mass dwarf galaxies~\cite{Nadler:2019zrb}.  This type of constraint must be considered less robust for now than the CMB constraints, since it relies on modeling non-linear regime processes and is subject to uncertainties in interpreting dwarf galaxy observations.  However since these limits may be put on a firmer footing in the future, we include the limits on ($\alpha,m_\phi,m_X$) using the constraints of ~\cite{DES:2020fxi} in Fig.~\ref{fig:alphamxAll}, for reference. 

\subsection{\label{sec:clouds}Milky Way Gas Clouds}
The limits on DM-baryon interactions from Milky Way gas clouds given in \cite{Wadekar:2019mpc} -- based on demanding that the heating/cooling rate of robust gas clouds due to scattering with DM particles not exceed the observed value -- were derived assuming Born approximation. However Born approximation exaggerates the DM-nucleus cross sections for nuclei heavier than H even more than for He, and thus the analysis of \cite{Wadekar:2019mpc} needs to be redone. The procedure to find $\sigma_0$ is similar to that discussed above for the CMB constraints and a relationship analogous to Eq.~\eqref{eq:sigHsig0} can be obtained, now including contributions of nuclei heavier than He on the LHS of the inequality. In the non-perturbative regime we are considering, only H and He scattering make a significant contribution to heating/cooling, unlike in the analysis of \cite{Wadekar:2019mpc} based on Born approximation. For an attractive interaction, the exact non-pertrubative treatment leads to the allowed region shown in Fig.~\ref{fig:sigsig}, and the limits in the $\alpha-m_X$ plane shown in Fig.~\ref{fig:alphamxAll} (left). 

\section{\label{sec:results} Combined Limits on the Dark Matter-Nucleon Interaction}

\subsection{Attractive Interaction}
\label{sec:attractiveresults}
Combining the results from the previous section, we present our final limits on $(\alpha,m_X)$ and $(\sigma_p,m_X)$ for $m_\phi= 1$ GeV and $R_0=1$ fm attractive interaction. The method of re-interpretation and analysis of experiments for repulsive interaction is the same and will be discussed second.

\subsubsection{\label{sec:alphcomb}Limits on Yukawa parameter $\alpha$}
Figure~\ref{fig:alphamxAll} (left panel) shows the allowed and excluded regions of ($\alpha,m_X$), for $m_\phi = 1 \text{\,GeV}$ and $R_0=1$ fm, for an attractive interaction, applying constraints from XQC,  CRESST, CMB, Lyman-$\alpha$ and Milky Way satellites, and the astrophysical limits from gas cloud heating.   The dashed lines indicate where $b=1$, above which the interaction is strong enough that the Born approximation breaks down. Evidently, in much of the parameter space of interest we cannot trust Born approximation, and in particular we cannot use the Born scaling~\eqref{eq:Ascaling} to draw our exclusion region, especially in the large coupling region probed by XQC and the CMB.  The CRESST limit is almost always in the $b\lesssim 1$ region so we expect less deviation from Born approximation and no resonances, except for the narrow anti-resonance region in which CRESST has sensitivity for $\alpha \gtrsim 0.1$ for an attractive interaction.  

There can be gaps in the exclusion from a single experiment due to resonant behavior of the cross section, such as for the CMB+Lyman-$\alpha$ constraints in Fig.~\ref{fig:alphamxAll} left. Moreover the exact positions of the gaps and boundaries move as the nuclear wave function and range of the Yukawa potential $m_\phi^{-1}$ are changed, so the positions of the boundaries should be considered blurry.  Such gaps are generally better overcome by considering multiple experiments with different target mass number $A$, rather than trying to improve the sensitivity for the same experiment, since  due to the non-trivial $A$-scaling in the resonant regime, different targets leave different gaps in the parameter space and the allowed region for one may be excluded for another.  

\subsubsection{\label{sec:xcnlims}Limits on $\sigma_p$}
\begin{figure}
\centering 
\includegraphics[width=1.0\textwidth]{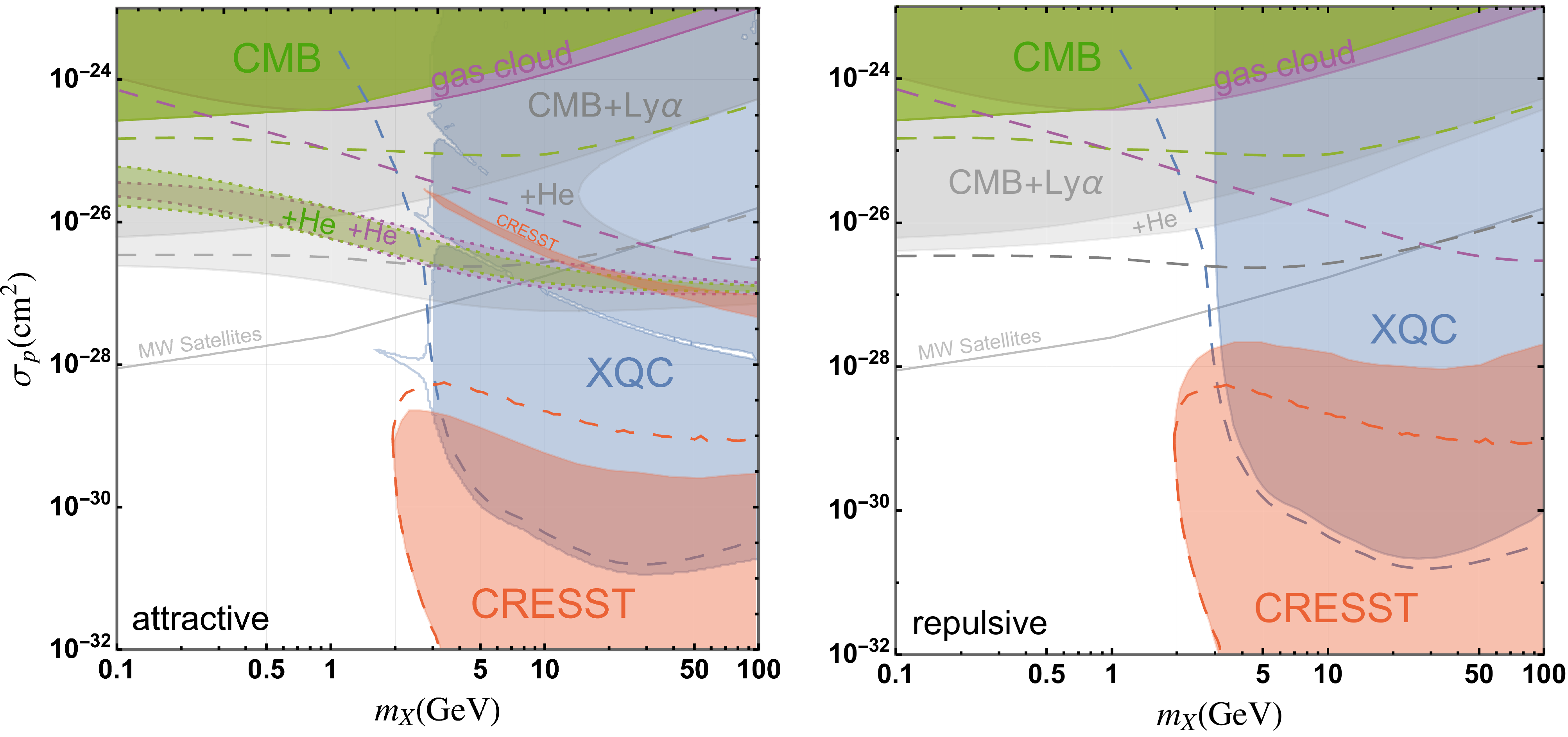}
\caption{\label{fig:sigmxAll} Exclusion region in ($\sigma_p$,$m_X$) space from various constraints for attractive (left panel) and repulsive (right panel) interactions, for $m_\phi=1 \text{\,GeV}$. Above $m_X \approx 3$ GeV the strongest limits come from direct detection experiments, XQC in blue and CRESST in red. For low masses, the strongest unambiguous limit is from CMB constraints on structure formation, shown as the green region labeled "CMB".  Astrophysical limits from gas cloud cooling, shown in purple, provide the strongest constraint in the 1-3 GeV range. For anattractive interaction, the green and purple bands surrounded by dotted lines and labeled "+He" are actually not fully excluded in the non-perturbative treatment if larger coupling $\alpha$ is allowed as discussed in Sec.~\ref{sec:CMB}, but if a theoretical limit of $\alpha < \mathcal{O}(1)$ is imposed these bands are excluded. For repulsive interaction, the CMB (green) and gas cloud (purple) limits cannot place any constraint on the ($\alpha,m_X$)parameter space, including DM-He scattering or not. So we only plot the unambiguous DM-proton scattering limits without re-interpretation. The gray regions labeled "CMB+Ly$\alpha$","+He"  and gray line  for "MW Satellites" are based on constraints which may not be robust and are included just for reference, as discussed in Sec.~\ref{sec:CMBplus}. The heavy dashed lines in the corresponding color show the previous Born-approximation-based limits. (The original CMB and Lyman-$\alpha$ limits (green and gray, dashed) have been modified slightly by changing the numerical factor in eq.~\eqref{bornXu} for He from 4 to 16 to include both proton and neutron scattering, as appropriate for isoscalar DM.) }
\end{figure}

Figure \ref{fig:sigmxAll} (left) shows the updated 
DM-proton cross-section limits from our analysis of XQC, CRESST and gas clouds, as well as the cosmology-based constraints.  The previous state-of-the-art limits are shown as dashed lines from XQC~\cite{Mahdawi:2017cxz,Mahdawi:2018euy}, CRESST~\cite{Mahdawi:2018euy}, CMB and CMB+Ly$\alpha$~\cite{Xu:2018efh} and gas clouds~\cite{Wadekar:2019mpc}.The derivation of our limits are discussed below in turn.

Starting with XQC, in~\cite{Mahdawi:2017cxz,Mahdawi:2018euy} it was assessed based on Born approximation that XQC had sensitivity to proton cross sections $\gtrsim 1$ mb for $m_X \leq 3 \text{\,GeV}$, via multiple scattering of a single DM particle in the detector allowing sufficient total energy deposit to meet the threshold condition.  (The relevance of multiple scattering for low masses was first pointed out in~\cite{Erickcek:2007jv}.)  However this apparent sensitivity disappears in our more accurate treatment, as can be understood from Fig.~\ref{fig:mxcombined}.  The salient points are that:
\begin{enumerate}
\item The XQC upper limit on $\sigma_{28}$ becomes weaker and weaker for lighter DM, because a larger and larger number of multiple scatterings in the detector are needed to produce a total energy deposit above threshold. 
\item For light DM requiring multiple scattering, e.g., $m_X= 2.9$ GeV shown in the left panel of Fig.~\ref{fig:mxcombined}, $\sigma_{28}^{\rm Born}>\sigma_{28}^{\rm Exact}$ for $\alpha> 0.2$.  Therefore, in this regime Born approximation exaggerates the ability to exclude.   In the narrow region around $\alpha \approx 0.1$, the opposite is true, resulting in an actually-excluded "peninsula" around $\sigma_p \approx 10^{-28}\, {\rm cm}^2$, which is not evident from the Born approximation. If we calculate to a larger $\alpha>10$, allowed $\alpha$ for $\sigma_{28}$ near a DM-proton anti-resonance could potentially produce $\sigma_p$ values in the "peninsula", reducing its size. However the position and extent of this excluded peninsula is not in any case well-determined, due to nuclear wave function sensitivity and our crude treatment of the velocity dependence for that part of the analysis, as discussed in Sec.~\ref{sec:reinterpretXQC},.
\item At higher mass, e.g., $m_X= 10 $ GeV shown in the right panel of Fig.~\ref{fig:mxcombined}, the more stringent limit on $\sigma_{28}$ requires $\alpha$ to be so small it is barely out of the Born regime and the treatment of~\cite{Mahdawi:2017cxz,Mahdawi:2018euy} is fairly accurate.  However since the exact $\sigma_{28}$ is larger than the Born value for $\alpha$ in this regime, the XQC limit is strengthened when the exact treatment is used. Allowing higher $\alpha>10$ will not make any difference because all $\alpha>10$ are excluded due to the strong constraint on $\sigma_{28}$: no anti-resonance can make $\sigma_{28}$ small enough to evade the limit.
\end{enumerate}

For CRESST, comparing the new limits given by the shaded regions to the old dashed limits in Fig.~\ref{fig:sigmxAll}, we see that  the true sensitivity region is generally reduced compared to what the perturbative analysis indicated, except for a sliver which appears at higher cross section.  The loss in CRESST's sensitivity to cross-sections $\sim 10^{-29} \, {\rm cm}^2$ falls in the XQC-excluded region, except for $2 \text{\,GeV} < m_X < 3 \text{\,GeV}$. 

According to our exact results, the limits from CMB alone (green solid in Fig.~\ref{fig:sigmxAll}) are considerably weaker than given by~\cite{Xu:2018efh} (green dashed in Fig.~\ref{fig:sigmxAll}). This is because \cite{Xu:2018efh} assumes a $4(\mu_{He}/\mu_H)^2$ times larger cross-section on He than on H, whereas in fact the He cross section is negligible for $\alpha$ large enough to produce a $\sigma_p$ at the limit. The green dotted "+He" region appears to exclude a range of cross-sections allowed by the Born approximation analysis (the old dashed limit). In this region, Born approximation \emph{underestimates} the true He contribution which is larger than Born due to a DM-He resonance. However, allowing a larger $\alpha$ up to $\sim$ 30 enables $\sigma_p$ to take these values due to an allowed H anti-resonance as shown in Fig.~\ref{fig:sigsig}. The gap between the confidently excluded solid region and the ambiguous dotted region is where the DM-He cross section is small and He contribution is negligible. 
The situation is the same for the gas cloud limit as the CMB limits, with just minor differences in the shapes of the exclusion regions and similar potential for large-$\alpha$ anti-resonance to produce non-excluded $\sigma_p$ in this range. 

If the CMB+Ly$\alpha$ limits can be used, the light gray region due to He scattering will only shrink a bit and not disappear completely if larger $\alpha$ are allowed, simply because the limit from Ly-$\alpha$ is stronger and the allowed $\alpha$ range is much smaller in the $(\alpha,m_X)$ plane. A large gap in the limits appears for $m_X \geq \,10  \text{\,GeV}$ due to He anti-resonance, which is, however, closed by XQC. Comparing to the limit obtained with only proton scattering, including Helium results in a stronger limit in general except in the gap caused by the small DM-He cross section near He anti-resonance, and allowing a larger $\alpha$ could reduce or eliminate the contribution of He.

If reliable, the Milky Way satellites limit for DM-proton scattering~\cite{DES:2020fxi} would be stronger than the CMB+Ly$\alpha$ limit, for $m_X \lesssim 10\text{\,GeV}$.  It is weaker for heavier DM when He scattering is included, although it would close part of the gap in the CMB+Ly$\alpha$ limit.  The underlying source of the satellites limit is inhibition of small scale structure formation from DM-baryon interactions, similarly to the CMB limits.  The limit relies on the validity of the mapping from linear-regime structure to as-observed dwarf galaxies used in~\cite{DES:2020fxi}, based on LCDM simulations which may not adequately describe interacting DM among other issues so its robustness needs further investigation.    

\subsection{\label{sec:repulsive}{Repulsive Interaction}}

Since the sign of the DM-$\phi$ coupling is unknown, in this subsection we repeat the analysis for a repulsive interaction with $\alpha \rightarrow -\alpha$ in Eq.~\eqref{eq:yukawa}. A repulsive potential does not allow the formation of bound states, so there is no resonance or anti-resonance for the cross section. At low energy the scattering is still s-wave dominated, and $\sigma \sim v^0$, i.e., no special velocity dependence. However, as for the attractive interaction, Born approximation breaks down when $\alpha$ (or the dimensionless parameter b) is large. This is clearly shown in Fig.~\ref{fig:ptextborn}. Most of the non-perturbative effects appearing in the analysis for attractive interaction still apply for the repulsive case. In particular, the non-trivial A dependence still persists and re-interpretation of constraints involving A>1 is needed.

The maximum value of the cross section that can be achieved with a repulsive interaction is also different from the attractive case. For an attractive interaction, the maximum cross section is achieved at the s-wave resonance, where the phase shift is close to $\pi /2$ and the cross section saturates the unitarity bound, as shown in Eq.~\eqref{eq:sigma_res}. The unitarity bound (Eq.~\eqref{eq:hi_v}) is 
\begin{equation}
    	\sigma = 4 \pi / (\mu v)^2.
\end{equation}
For Galactic dark matter with $v\gtrsim$ 100 km/s and GeV-scale mass, $\sigma \lesssim 10^{-21} \rm{\,cm^3}$ on resonance. The CMB limit as shown in Fig.~\ref{fig:experiments} is strong enough to constrain such resonant values. However, for the repulsive case, without the resonance, the cross section is bounded by the range of the interaction. When the radius of the nucleus is much larger than the Compton wavelength of the mediator, $r_A \gg 1/m_\phi$, the limit is given by the geometrical size of the nucleus, $\sigma_A \lesssim 4\pi r_A^2$. This is seen in Ref.~\cite{Digman:2019}, where a repulsive finite square well potential is used.  The corresponding limit in the extended Yukawa potential~\eqref{eq:Vball} is $r_A \, m_\phi \rightarrow \infty$. For a proton, $r_{\rm{H}} \sim \rm{\,fm}$ and the cross section caps at $\sigma_{\rm{H}} \sim 10^{-25} \rm{\,cm^2}$.  With the maximum possible $\sigma_{\rm{H}}$ being smaller than the CMB limits, the repulsive parameter space is unconstrained by these limits. This roughly applies for our benchmark $m_\phi \sim$ GeV. For a lighter mediator, the cross section is determined by $1/m_\phi$ rather than $r_A$ and could be greater than the CMB limit.  ~\cite{Digman:2019} also discussed another possibility for the cross section to exceed the size of nucleus, so that the CMB limit can be relevant: when the dark matter particle itself is not point-like, in which case the cross section can be as large as the size of the DM particle instead of the nucleus. A follow-up experimental analysis ~\cite{Cappiello:2020lbk} places corresponding limits on non-point-like dark matter.

\subsubsection{\label{sec:alphcomb}Limits on Yukawa parameter $\alpha$}

Figure~\ref{fig:alphamxAll} (right) shows the constraints in the $(\alpha,m_X)$ plane for a repulsive interaction. There are several differences from its attractive counterpart. The XQC boundary gets smoothed out and no gap is formed, because of the lack of (anti-)resonance. The cross section depends monotonically on the underlying parameters so no gaps appear in the excluded domain. The upper reach of CRESST is increased because the repulsive interaction has a generally smaller cross section so the overburden produces less shielding 
the anti-resonance excluded band at large alpha also disappears. The CMB limit disappears entirely due to the inability to saturate the cross section limit for the repulsive interaction, as discussed previously. Actually, detailed calculation shows that for $m_\phi = 1 \rm{\,GeV}$, the CMB limit is only sensitive in the repulsive case to $\alpha \gtrsim 10^{3}$ for $m_X = 1 \rm{\,GeV}$ and $\alpha \gtrsim 10^{8}$ for $m_X = 10 \rm{\,GeV}$, and the contribution from DM-He scattering is ignorable. The gas cloud limit is similar to the CMB one and also does not contribute to the constraint on ($\alpha,m_X$). The CMB+Lyman-$\alpha$ limit, if reliable, would be much stronger than the CMB-only limit and including DM-He scattering would improve the bound; however there is no gap like the one seen in the left panel for the attractive case.

\subsubsection{\label{sec:xcnlims}Limits on $\sigma_p$}
Figure~\ref{fig:sigmxAll} (right) shows the exclusion region for $(\sigma_p,m_X)$ for a repulsive interaction. The boundary of the XQC excluded region is smoothed and the upper reach of CRESST is increased. The green CMB limit and cyan gas cloud limit are the original constraints including only proton scattering and are what we can safely trust without re-interpretation. In general, the repulsive interaction cannot achieve these cross sections, or even if it does, with extremely large $\alpha$, the contribution of He and heavier nuclei can be ignored because they have similar geometrical size as the proton but are significantly less abundant. If the CMB+Lyman-$\alpha$ limit can be used, including He does improve the constraint, albeit not as much as with the (invalid) Born approximation prescription, because $\sigma_{\rm {He}}/ \sigma_{\rm{H}}$ is much smaller than the Born scaling Eq.~\eqref{eq:Ascaling}.

\subsection{\label{sec:mphi}Dependence on Mediator Mass}
We adopted $m_\phi = 1 \text{\,GeV}$ for our analysis, to be concrete and because for sexaquark DM the dominant interaction with nucleons is through exchange of the flavor-singlet combination of $\omega-\phi$ vector mesons, whose mass is in this range. For heavier-than-GeV $m_\phi$, and for heavy target nuclei such as Si in XQC and the majority of nuclei in CRESST, the condition $r_A \gg 1/m_\phi$ is well satisfied and the scattering potential can be approximated by a uniform spherical well with radius $r_A$ and depth $V_0 \propto \alpha / m_\phi^2$, as indicated in eq.~\eqref{eq:Vballinside}. In this case, the scattering cross section for a given nucleus  is only a function of $\alpha / m_\phi^2$. As a result, our exclusion region in Fig.~\ref{fig:alphamxAll} is the same for a different $m_\phi$, except for a re-scaling of the $\alpha$-axis by the factor $\alpha \sim  m_\phi^2$.  Returning to the limits on the scattering cross section for a given nucleus undoes this rescaling, with the result that the limits in Fig.~\ref{fig:sigmxAll} remain the same for any $m_\phi > 1$ GeV.

For lighter nuclei such as H and He and for lighter $m_\phi \lesssim $ 100 MeV, we have $r_A \lesssim 1/m_\phi$ and the potential deviates from being a spherical well. The cross sections cannot be obtained by any simple re-scaling of the $m_\phi \sim$ GeV result and need to be re-calculated. However, there are still some simplifications for light $m_\phi$: 
\begin{enumerate}
    \item For $m_\phi \lesssim 10$ MeV, we have $r_A \ll 1/m_\phi$ or $c/b \ll 1$ and the point Yukawa potential Eq.~\eqref{eq:yukawa} is enough. The size of the nucleus does not matter anymore. 
    \item For $m_\phi \sim$ (0.1-1) MeV the Born approximation Eq.~\eqref{eq:sm-cxtotborn} turns out to be quite accurate for $\sigma_p \lesssim 10^{-21} \rm{\,\,cm^2}$. However the oft-used "Born scaling" to trivially relate cross sections for different $A$ using Eq.~\eqref{eq:Ascaling} is still wrong in this case, since we do not have $\mu v \ll m_\phi$ or $ab \ll 1$ now. The full Born approximation Eq.~\eqref{eq:sm-cxtotborn} should be used.
    \item For $m_\phi \lesssim$ 0.1 MeV, we have $ab \gg 1$ which enters into the classical regime and there is no resonance, see Fig.~\ref{fig:2Dab}. The classical problem has been solved and a fitting function was given in Ref.~\cite{Khrapak:2004ieee,Loeb:2010gj}:
\begin{equation}
\label{eq:sigma_classical}
\sigma_T \approx \begin{cases}\frac{4 \pi}{m_{\phi}^{2}} \beta^{2} \ln \left(1+\beta^{-1}\right), & \beta \lesssim 0.1 \\ \frac{8 \pi}{m_{\phi}^{2}} \beta^{2} /\left(1+1.5 \beta^{1.65}\right), & 0.1 \lesssim \beta \lesssim 10^{3} \\ \frac{\pi}{m_{\phi}^{2}}\left(\ln \beta+1-\frac{1}{2} \ln ^{-1} \beta\right)^{2}, & \beta \geqslant 10^{3}\end{cases}
\end{equation}
where $\beta=\alpha m_{\phi} /\left(\mu v^{2}\right)$.
Notice here many partial waves contribute and the momentum transfer cross section has to be used to ensure convergence.
\end{enumerate}
We include Fig. \ref{fig:smallmphi} in the appendix to show the limits on ($\sigma_p,m_X$) for $m_\phi = (1, 10, 100)$ MeV for completeness. Cross sections for $m_\phi \lesssim 1$ MeV can be calculated analytically using Born approximation or classical fitting function as described above.

\subsection{\label{sec:digman} The interpretation of a DM-baryon cross section exceeding $10^{-25}\,{\rm cm}^2$}

Regardless of whether the DM is pointlike or composite, the DM-baryon cross section can exceed $10^{-25}\,{\rm cm}^2$,  contrary to statements in the literature.\footnote{
E.g., Ref.~\cite{Digman:2019}  in the conclusion: 
"(3) For $\sigma_{\chi N} > 10^{-25}\, {\rm cm}^2$, dark matter cannot be pointlike. Contact interactions cannot achieve cross sections larger than the geometric cross section $\sigma_{\chi A} = 4 \pi r_A^2$, and simple light mediators are strongly ruled out. Dark matter with cross sections in this range must be composite." and Ref.~ \cite{Cappiello:2020lbk}, e.g.. in the abstract:  "Recently, it was shown theoretically that the scattering cross section for $m_\chi \gtrsim 1$ GeV pointlike dark matter with a nucleus cannot be significantly larger than the geometric cross section of the nucleus. This realization closes the parameter space for pointlike strongly interacting dark matter."}  
This can be seen both in our explicit calculations and from partial wave unitarity.  Unitarity gives the maximum s-wave cross section (Eq.~ \eqref{eq:sigmadelta}) 
\begin{equation*}
    \sigma_{\rm peak} = \frac{4 \pi}{\mu^2 \, v^2} = 4.9\times 10^{-21}\,{\rm cm}^2 \, \left(\frac{\rm GeV}{\mu}\frac{10^{-3} c}{v}\right)^2~,
\end{equation*}
where $\mu$ is the reduced mass and $v$ is the relative velocity.  The unitarity-limit cross section is reached when the parameters of the DM-nucleus or DM-nucleon potential are such that the system has a zero energy bound state; this can occur whether or not the DM particle is point-like. Note that the s-wave unitarity limit 
(Eq.~\eqref{eq:sigmadelta}) is the same whether the particles scattering are pointlike or extended.  Figure~\ref{fig:ptextborn} shows $\sigma_{p}$ for a point-like DM particle and both a point and extended nucleon as a function of $\alpha$, for $(m_X,\, m_\phi) = (2 m_p,\, 1\, {\rm GeV})$ and relative velocity 300 km/s.  The peak cross sections occur at different values of $\alpha$ for the point-like and extended nucleon cases, but in both instances they reach the same unitarity limit value given above: $10^{-20.3}\,{\rm cm}^2$ for $v=$ 300 km/s.  (The apparent difference in peak heights in Fig.~\ref{fig:ptextborn} is due to the discretization of the plotting function.)

Whether or not the unitarity bound can be saturated depends on whether the interaction is repulsive or attractive.  The existence of resonances and the possibility of near-saturation of unitarity is a very general feature of attractive potentials.  In the case of a Yukawa sourced by a nucleus, the particular coupling strength giving rise to a resonance depends on the mass of the mediator and the size and shape of the source. For a fixed coupling strength and mediator mass, how close the cross section comes to saturating unitarity depends on the nuclear size $A$.  Using another nuclear wave function than a simple sphere would also shift the parameter values of the resonance.   However the existence of a near-resonance is generic, as long as a near-zero-energy bound state exists for some $A$ given the fundamental parameters.  This leads to large cross sections with the velocity dependence discussed in Sec.~\ref{app:vdep}. However if the potential is repulsive it does not admit zero-energy bound states and unitarity is not saturated.  The maximum cross-section for pointlike DM in the repulsive case is therefore more limited, as discussed in Sec.~\ref{sec:repulsive}.  With a mediator no lighter than a pion, the cross section can only reach $10^{-24.1}\,{\rm cm}^2$.  

The possibility of using the magnitude of the DM-nucleus cross section as a diagnostic of dark sector particles having extended structure is certainly tantalizing.  If the distribution of $\mu v$ for the DM were known and the cross section were established to exceed the s-wave unitarity limit, that would be evidence of multiple partial waves contributing simultaneously and suggest the de Broglie wavelength $1/ \mu v$ is smaller than the length scale in the scattering system. This scale can be the size of the nucleus, the size of the DM particle, or the Compton wave length of the mediator $1/m_\phi$, whichever is the largest. A detailed analysis would be required to decide. Measuring the cross section for a variety of nuclei $A$ would be a valuable diagnostic.  Constraints on light mediator candidates and $m_\phi$ could also help narrow down the possibilities~\cite{Digman:2019}.

\section{\label{sec:conclusion} Final Joint Limits and Conclusions}
We have shown that the pioneering and state-of-the-art analyses of direct detection and CMB and astrophysical constraints are not generally valid due to inappropriate use of Born approximation to relate the cross sections for nuclei of different $A$ to that of protons.  Furthermore earlier analyses did not properly take into account the finite size of nuclei.

DM-baryon elastic scattering via a massive mediator in general exhibits resonance behavior if the interaction is attractive and can depart significantly from the Born approximation result even for repulsive interactions.  Thus DM interactions with baryons must be analysed by exact numerical solution of the Schrödinger equation in a substantial portion of interesting parameter space.  For example, GeV-range dark matter with an attractive Yukawa interaction lies in the non-perturbative resonant regime for the XQC experiment with $A=28$, even for Yukawa coupling strength as low as $\alpha = 0.02$.  

In the resonant regime, it is non-trivial to interpret results of direct detection experiments and other constraints from observations, particularly when $A>1 $ nuclei are used as targets.  Universal limits on the DM-nucleon cross section cannot be directly obtained as long as it is the DM-$A$ cross section that is actually experimentally constrained, since there is a non-trivial, model-dependent relationship between $\sigma_p$ and $\sigma_A$. For example, XQC using a silicon detector leaves a large part of ($\sigma_p,m_X$) parameter space for $m_X \lesssim 3 \text{\,GeV}$ allowed, rather than being excluded as concluded previously based on naive use of Born approximation.  

Analyses of CMB constraints also need modification, due to the non-trivial relation between DM-He and DM-p cross sections.  When this is taken into account, the CMB limits are weakened because large $\sigma_p$ does not imply large $\sigma_{\rm He}$ or vice-versa.  If the CMB+Ly-$\alpha$ constraints are valid (which is not yet clear, given current limited understanding of patchy reionization \cite{Hui:2016ltb,vandenBosch+Shattering19}), they would strengthen the limits relative to CMB-only limits but still leave a gap in the exclusion region for relatively large coupling, due to anti-resonance behavior which sharply reduces the DM-He cross-section for DM mass above 10-15 GeV.  At the same time, the correct treatment strengthens CMB+Ly-$\alpha$ limits for lower masses and smaller $\sigma_p$, due to the non-perturbative enhancement of  $\sigma_{\rm He}$ in this parameter regime.  A further strengthening of the limits would be possible if the Milky Way satellites analysis proves robust, as shown in Fig.~\ref{fig:sigmxAll}.

\begin{figure}
\centering 
\includegraphics[width=0.6\textwidth]{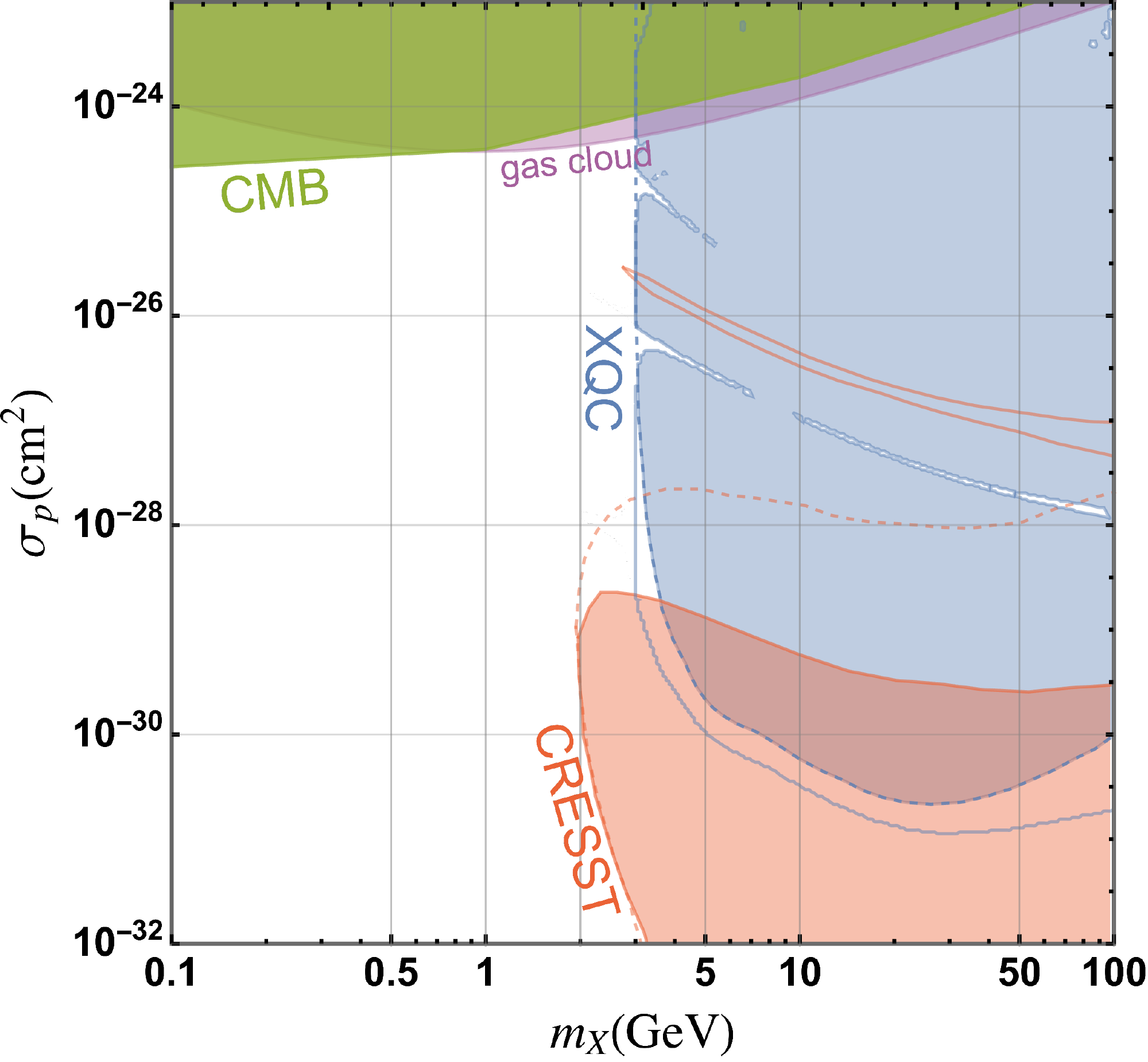}
\caption{\label{fig:combined} The final exclusion regions for $\sigma_p$ as a function of $m_X$, for $m_\phi = 1$ GeV, adopting the more conservative assumption on the sign of the interaction for each parameter combination and 
not including CMB+Ly-$\alpha$ and Milky Way satellites constraints which may not be robust.
The XQC limits are more conservative (weaker) for the repulsive case, while the CRESST exclusion region is a mix.  The island at large cross-section which is excluded if the interaction is attractive, due to the anti-resonance in that case, is allowed for a repulsive interaction and thus cannot be excluded in the absence of knowledge of the sign.  Similarly, the conservative choice for the lower excluded region is for the attractive sign since it extends to higher cross section if the interaction is repulsive. For $m_X \lesssim$ 2 GeV and $\sigma_p \lesssim {\rm few}\times 10^{-25} \mathrm{\,\,cm}^2$ the parameter space remains unconstrained.  Measuring the thermalization efficiency of XQC to determine the accurate $m_X$ value of its lower reach is very important; a similar uncertainty applies to CRESST.  } 
\end{figure}

Our summary results on the excluded regions for the Yukawa coupling and DM-nucleon cross-section are shown in  Figs.~\ref{fig:alphamxAll} and \ref{fig:sigmxAll} for the attractive and repulsive cases separately.  Figure~\ref{fig:combined} combines these results and gives the final constraints which can be placed today on $(\sigma_p,m_X)$ using XQC, CRESST, CMB and Gas Clouds, if the sign of the Yukawa coupling is unknown. Even the step of combining the limits presents a nuance.  A plot could usefully aim to address two different questions: (1) Can the data from a given experiment exclude a given point in $(\sigma_p,m_X)$, independent of the sign of the interaction? or (2) Is a given point in $(\sigma_p,m_X)$ inconsistent with all experiments sensitive to that $m_X$?  The excluded region of type (1) for a particular experiment is the intersection of the excluded regions in the left and right panels of Fig.~\ref{fig:sigmxAll} for the same experiment. These experiment-by-experiment fully-excluded regions are shown using solid colors in Fig.~\ref{fig:combined}, for each of the robust constraints:  CMB, gas clouds, XQC and CRESST. 

Additionally, Fig.~\ref{fig:combined} shows how the exclusion region increases for each experiment if the sign were {\it a priori} known, using solid for an attractive interaction and dashed for repulsive. The intersection of the XQC and CRESST solid regions is even more robustly excluded than the non-overlapping regions, due to having independent experiments giving the same conclusion.  However the wedge bounded by blue solid and blue dashed lines, and red solid and red dashed lines, (containing $(\sigma_p,\, m_X) \approx (10^{-28}\,{\rm cm^2}, 3\, {\rm GeV})$), is not excluded by both experiments because it is excluded by XQC only if the interaction is attractive, and by CRESST only if it is repulsive.  

The interesting parameter space for sexaquark DM ($m_X \lesssim 2$ GeV) remains largely unconstrained, requiring only $\sigma_p \lesssim 10^{-24.5} \, {\rm cm}^2$.  The constraints on the Yukawa coupling parameter are quite limited for DM mass below XQC  sensitivity, $m_X \lesssim 3$ GeV,  with only small ranges of $\alpha$ being excluded. If Ly-$\alpha$ measurements 
can be used to constrain the growth of small scale structure, $\alpha \gtrsim 0.3$ could be excluded for attractive interaction and $m_X \approx 2$ GeV, leading to an upper limit on the cross section in the mb range. That would be constraining but not challenging for sexaquark DM.  The constraints are stronger for heavier DM, where the energy deposit per collision is larger and the direct detection experiments XQC and CRESST are more sensitive.  As pointed out in~\cite{Mahdawi:2018euy,Farrar:2020}, it is crucial to measure the thermalization efficiency of semiconductors to small nuclear recoils before experiments like XQC and CRESST can be interpreted for lower DM masses.  It should also be noted that if the thermalization efficiency is smaller than the 0.01 adopted here, as may be the case, the limits should be weaker than indicated and those indicated in the plot would not be conservative.

Recently, in ref.~\cite{Neufeld:2019xes} the authors performed experiments with 27 different atomic nuclei and reported limits on $\sigma_{DM-A}$ for each of them, extending the limits of ref.~\cite{Neufeld:2018slx} derived by consideration of the atmosphere of DM particles surrounding the Earth and its impact on the evaporation of cryogens and drag on the Hubble Space Telescope.  Due to the complexity of the analysis, we leave consideration of ~\cite{Neufeld:2018slx,Neufeld:2019xes} for elsewhere.  However the combination of the lower mass reach and large coverage in $A$ of the constraints of~\cite{Neufeld:2018slx,Neufeld:2019xes}, offers promise that these will be a valuable addition to the constraints on DM interactions with hadrons, for DM in the GeV mass range. \footnote{Since the preliminary version of the present paper  (arXiv:2101.00142v1~\cite{Xu:2020qjk_v1}) was posted and now, we have finished analysing the dewar experiment results.  That work is reported in~\cite{Xu:2021dewar}.  It includes a combined plot comparing with the limits derived here.}

\acknowledgments
We thank M.~S.~Mahdawi for discussions,  and for the use of his code developed for the studies reported in~\cite{Mahdawi:2017cxz, Mahdawi:2018euy},  and thank J.~F.~Beacom and C.~V.~Cappiello for helpful comments and discussions about the manuscript. XX received support from a James Arthur Graduate Fellowship; the research of GRF has been supported in part by the Simons Foundation and by NSF-PHY-1517319 and NSF-PHY-2013199.


%
%
%
%
%

\bibliography{xxcbib}
\bibliographystyle{JHEP}
\newpage

\begin{center}

\large{\bf APPENDICES}
\end{center}
\vspace{-0.2in}

\appendix
\section{\label{sec:numerical}Numerical Methods}

To obtain the total elastic scattering cross section for DM-baryon scattering in the non-relativistic regime, we need to solve the Schrödinger equation exactly with the appropriate point or extended Yukawa potential and perform a partial wave analysis. Two approaches are discussed below; which is superior depends on regime. 
\subsection{Schrödinger Equation Method} In the first approach, we follow the method described in~\cite{Buckley:2010, Tulin:2013teo}. Although that work addresses the collision between DM particles, the mathematical problem is exactly the same. The radial Schrödinger equation in this problem is
\begin{equation}
\label{eq:schdim}
\left \{  {\frac{1}{r^2}} {\frac{d}{dr}}\left(r^2{\frac{d}{dr}}\right) +k^2-{\frac{l(l+1)}{r^2}}-2\mu V(r) \right \} R_l(r)=0
\end{equation}
with $r$ the distance between two scatterers.  $\mu$ is the reduced mass of DM and nuclues,  $\mu=m_X m_A/(m_X+m_A)$, and $R_l(r)$ is the radial component of the wave function for partial waves $l=0,1,2..$. The phase shifts $\delta_l$ parametrize the asymptotic behavior of the wave function when $r$ goes to infinity
\begin{equation}
\label{eq:match1}
\lim_{r \rightarrow \infty}R_l(r) \propto \sin(kr-\pi l /2 +\delta_l),
\end{equation}
where $k=\mu v$, and $v$ is the relative velocity of the two particles. The differential and total cross section in terms of the phase shifts are given by
\begin{equation}
\label{eq:diffsigmadelta}
\frac{d\sigma}{d\Omega} =\frac{1}{k^2} \left | \sum_{l=1}^{\infty} (2l+1)e^{i\delta_l}P_l(\cos\theta)\sin\delta_l\right |^2,
\end{equation}
\begin{equation}
\label{eq:sigmadelta}
\sigma =\frac{4\pi}{k^2}\sum_{l=0}^{\infty }(2l+1)\sin^2(\delta_l).
\end{equation}
In the DM detection literature the momentum transfer cross section $\sigma_T$ is often used:
\begin{equation}
\label{eq:sigmaTdelta}
\sigma_T =2\pi\int \frac{d\sigma}{d\Omega}(1-\cos{\theta})d\theta=\frac{4\pi}{k^2}\sum_{l=0}^{\infty }(l+1)\sin^2(\delta_{l+1}-\delta_{l}).
\end{equation}
We will see later that in the parameter space we are interested in, s-wave ($l=0$) scattering is dominant. So there is little difference between~\eqref{eq:sigmadelta} and~\eqref{eq:sigmaTdelta} and we will use the total cross section~\eqref{eq:sigmadelta} throughout this paper.
Now specializing to the Yukawa potential~\eqref{eq:yukawa} and changing variables to the dimensionless combination
\begin{equation}
\label{eq:changevariable}
\begin{split}
x &\equiv 2\mu \alpha r\,,
\qquad
u_l(x) \equiv rR_l(r) \,,
\\
a &\equiv \frac{v}{2\alpha} \,,
\quad \quad \quad \quad \quad
b \equiv \frac{2 \mu \alpha}{m_\phi} \,,
\end{split}
\end{equation}
the Schrödinger equation~\eqref{eq:schdim}  becomes  
\begin{equation}
\label{eq:schdimless}
\left \{   {\frac{d}{dx^2}} +a^2-{\frac{l(l+1)}{x^2}} - \tilde{V}(x) \right \} u_l(x)=0,
\end{equation}
where the dimensionless potential is
\begin{equation}
\label{eq:vdimlesspoint}
 \tilde{V}(x)=-{\frac{1}{x}}e^{-\frac{x}{b}}.
\end{equation}
In this notation the phase shift is given by
\begin{equation}
\label{eq:match2}
\lim_{x \rightarrow \infty}u_l(x) \propto \sin(ax-\pi l /2 +\delta_l).
\end{equation}
The cross sections are
\begin{equation}
\label{eq:diffsigmaab}
\frac{d\sigma}{d\Omega} =\frac{1}{a^2b^2m_\phi^2} \left | \sum_{l=1}^{\infty} (2l+1)e^{i\delta_l}P_l(\cos\theta)\sin\delta_l\right |^2
\end{equation}
\begin{equation}
\label{eq:cxtotSI}
\sigma=\frac{4\pi}{a^2 b^2 m_\phi^2}\sum_{l=0}^{\infty }(2l+1)\sin^2(\delta_{l})
\end{equation}
where $k=abm_\phi$. The phase shift $\delta_l$ and $\sigma m_\phi^2$ only depend on the dimensionless parameters $(a,b)$.  For reference, (first order) Born approximation gives
\begin{equation}
\label{eq:sm-cxdifborn}
\left( \frac{d\sigma}{d\Omega} \right)^{\rm{Born}}=\left( \frac{2 \mu \alpha}{m_\phi^2 + 4k^2\sin^2\frac{\theta}{2}} \right)^2
\end{equation}
\begin{equation}
\label{eq:sm-cxtotbornApx}
\sigma^{\rm{Born}} =\frac{16 \pi \mu^2 \alpha^2}{m_\phi^2(m_\phi^2+4k^2)}
=\frac{4\pi b^2}{m_\phi^2 (1+4 a^2 b^2)}.
\end{equation}
Eq.~\eqref{eq:schdimless} is a second order differential equation together with the initial condition $u_l(0)=1,u_l'(0)=0$ (for regularity of the wave function at $x=0$). It needs to be solved for $u_l(x)$ with $x\in [0,\infty)$, and $\delta_l$ is obtained from $u_l(\infty)$ as in~\eqref{eq:match2}. However, numerically the best we can do is to solve $u_l(x)$ for $x\in [x_i,x_m]$ with sufficiently small(large) but finite $x_i$($x_m$),  and match $\delta_l$ at $u_l(x_m)$ to achieve appropriate accuracy. The detailed numerical method for the point Yukawa potential is described below. For the extended Yukawa potential, a new scale "c" is introduced and a slight modification is needed.
 
\noindent(1) Since we are changing the initial point from zero to $x_i>0$, a different initial condition should be used. When $x_i$ is sufficiently small, such that the Yukawa and $a^2$ terms in eq.~\eqref{eq:schdimless} are sub-dominant compared to the angular momentum term, the solution is approximately $u_l(x) \propto x^{l+1}$. Ignoring the overall normalization of $u(x)$ since it is irrelevant for determining $\delta_l$, we impose the initial condition:
\begin{equation}
\label{eq:ic}
\begin{split}
u_l(x_i)&=1
\\
u'_l(x_i)&=\frac{l+1}{x_i}.
\end{split}
\end{equation}
$x_i$ should be small enough for eq.~\eqref{eq:ic} to work, and we choose the following condition based on trail and error in our parameter range,
\begin{equation}
\label{eq:inixi}
\begin{cases}
x_i \leq \frac{1}{10}\min{(1,\frac{1}{a},b)} &(\text{for} \quad l=0)
\\
\\
a^2 \leq \frac{1}{10} \frac{l(l+1)}{x_i^2} \quad   \frac{1}{x_i}e^{-\frac{x_i}{b}} \leq \frac{1}{10} \frac{l(l+1)}{x_i^2}  \quad  x_i\leq \frac{1}{10}x_{it}  &(\text{for} \quad l>0)
\end{cases}
\end{equation}
where $x_{it}$ is the smallest classical turning point. (Notice that for $l=0$ the angular momentum term disappears and a different condition must be used.) This is our initial guess for a small enough $x_i$.

\noindent(2) We solve eq.~\eqref{eq:schdimless} with the initial condition~\eqref{eq:ic} for $x\in [x_i,x_m]$. The end point $x_m$ is determined so that the Yukawa term is negligible compared to the $a^2$ term and the angular term. For any potential which is exactly zero for $x\geq x_m$, matching the solution at $x_m$ will give the exact $\delta_l$. The condition for $x_m$  is
\begin{equation}
\label{eq:inixm}
\begin{cases}
\frac{1}{x'_m}e^{-\frac{x'_m}{b}} \leq \frac{1}{10}a^2 \quad x_m\geq x'_m+5b   \quad\quad\quad\quad \quad\quad \quad\quad \quad\quad\quad&(\text{for} \quad l=0)
\\
\\
\frac{1}{x_m}e^{-\frac{x_m}{b}} \leq \frac{1}{10}a^2 \quad  \frac{1}{x_m}e^{-\frac{x_m}{b}} \leq \frac{1}{10} \frac{l(l+1)}{x_i^2}  \quad x_m\geq x_{mt}+5b  \quad\quad &(\text{for} \quad l>0)
\end{cases}
\end{equation}
where $x_{mt}$ is the largest classical turning point. The condition involving $x_{mt}$ is critical since in practice we find the phase shift only starts to converge to its asymptotic value after $x_{mt}$, where the wave function starts to oscillate like a sine function. The $5b$ is chosen from experience, justified by the fact that $b$ is the only relevant scale introduced by the Yukawa term. This will be our initial guess for a large enough $x_m$.

\noindent(3) At $x_m$ we match $u_l(x)$ to a new asymptotic form, different from eq.~\eqref{eq:match2}:
\begin{equation}
\label{eq:matchTulindimless}
\lim_{x \rightarrow \infty}u_l(x) \propto xe^{i \delta_l} \left( \cos{\delta_l}j_l(ax)-\sin{\delta_l}n_l(ax)  \right)
\end{equation}
following \cite{Tulin:2013teo}, where $j_l$ ($n_l$) is the spherical Bessel (Neumann) function. The corresponding dimensional condition for the original wave function is
\begin{equation}
\label{eq:matchTulindimful}
\lim_{r \rightarrow \infty}R_l(r) \propto  \cos{\delta_l}j_l(kr)-\sin{\delta_l}n_l(kr).
\end{equation}
 Inverting~\eqref{eq:matchTulindimless}, the phase shift is given by
\begin{equation}
\label{eq:phaseTulin}
\begin{split}
&\delta_l=\arctan \left[ {\frac{a x_m j'_l(a x_m)-\beta_l j_l (a x_m)}{a x_m n'_l(a x_m)-\beta_l n_l (a x_m)}} \right],
\\
&\text{where }\beta_l \equiv \frac{x_m u'_l(x_m)}{u_l(x_m)}-1.
\end{split}
\end{equation}

\noindent(4) Based on the initial guess of ($x_i$,$x_m$), we calculate $\delta_l$ according to~\eqref{eq:phaseTulin}. To check convergence, in other words to test if $x_i(x_m)$ is sufficiently small(large), we first fix $x_i$ and increase $x_m$ in units of $b$ (by the assumption that $b$ is the relevant scale for convergence) until $\delta_l$ converges at $1\%$, to obtain a ($x_i$,$\delta_l$) pair. Practically, we find $\delta_l$ will converge within $x_m+20b$. Then we decrease $x_i$ by half each time until $\delta_l$ converges at 1$\%$ with respect to $x_i$. Our experience shows convergence will be achieved within $2^{-20}x_i$. Finally we obtain $\delta_l$ for each ($a$,$b$,$l$).

\noindent(5) We calculate $\sigma$ by summing over $l$ in eq.~\eqref{eq:cxtotSI}. The truncation of the series, $l_{\mathrm{max}}$, is determined by the requirement that $l_{\mathrm{max}}$ contributes to $\sigma$ less than 1$\%$ of the sum of all smaller $l$, and $\delta_{l_{\mathrm{max}}}<$0.01. And we check this condition for 10 successive $l_{\mathrm{max}}$ to make sure we arrived at sufficiently large $l$ so that higher $l$ makes ignorable contribution.

\subsection{Phase Function Method}
An alternative way to calculate the phase shift $\delta_l$ instead of solving the Schrödinger equation, is the phase function method or variable phase method. See the book by Calogero~\cite{Calogero:PFM} for details. In our dimensionless parametrization, we define the differential equation obeyed by the phase function $\delta_l(x)$:
\begin{equation}
\label{eq:pfm1}
\delta'_l(x)=axe^{-\frac{x}{b}}\left[ \cos{\delta_l(x)j_l(ax)} - \sin{\delta_l(x)n_l(ax)} \right]^2
\end{equation}
with the boundary condition
\begin{equation}
\begin{split}
\lim_{x \rightarrow 0}&\delta_l(x) \rightarrow \frac{a^{2l+1}x^{2l+2}}{(2l+2)[(2l+1)!!]^2} \rightarrow 0
\\
\lim_{x \rightarrow \infty}&\delta_l(x) \rightarrow \delta_l
\end{split}.
\end{equation}
It is clear that the phase function got it is name since its asymptotic value gives the phase shift: $\delta_l(\infty) = \delta_l$. 
The advantage of the phase function method is that the phase equation is 1st order instead of the 2nd order Schrödinger equation. The trade-off is that the equation is now non-linear. In the form described by equation~\eqref{eq:pfm1}, the angular momentum term is eliminated, so we need to integrate to a larger cut-off point $x_m$ to achieve appropriate convergence for $\delta_l$. Overall, the phase function method turns out to be more efficient and has the ability to resolve a much smaller phase shift than the Schrödinger equation method, in certain regions of the parameter space. In obtaining our results for this paper, both methods have been used depending on the parameter regime. Our program is written in Mathematica.


\section{Approximate Treatment of Velocity Dependence of Extended Yukawa}
\label{app:vdep}
\subsection{Summary of key potentials}

\paragraph{Point Yukawa} The point Yukawa potential takes the form
\begin{equation}
\label{eq:pointYukawa}
\begin{split}
V(r)&=-\frac{\alpha}{r}e^{-\frac{r}{r_\phi}}
\\
\tilde{V}(x)&=-\frac{1}{x}e^{-\frac{x}{b}},
\end{split}
\end{equation}
where we defined
\begin{equation}
\label{eq:b}
b=2 \mu \alpha r_\phi = \frac{2 \mu \alpha}{m_\phi},
\end{equation}
\begin{equation}
\label{eq:x}
x=2 \mu \alpha r.
\end{equation}

\paragraph{Extended Yukawa} The extended Yukawa takes the form
\begin{equation}
\label{eq:Vball_appdix}
\begin{split}
V(r)&=-\frac{3 \alpha}{m_{\phi}^{2} r_0^3} \times
\begin{cases}
1-(1+m_\phi r_0) e^{-m_\phi r_0}\frac{\sinh{(m_\phi r)}}{m_\phi r} \quad &(r < r_0)
\\
\left[ m_\phi r_0 \cosh{(m_\phi r_0)} - \sinh{(m_\phi r_0)}  \right] \frac{e^{-m_{\phi} r}}{m_\phi r} \quad &(r\geq r_0) 
\end{cases}
\\
\tilde{V}(x)&=-3\left(\frac{b}{c}\right)^3 \times
\begin{cases}
\frac{1}{b}-(1+\frac{c}{b}) e^{-\frac{c}{b}}\frac{1}{x} \sinh{(\frac{x}{b})} \quad &(x < c)
\\
\left[ \frac{c}{b} \cosh{(\frac{c}{b})} - \sinh{(\frac{c}{b})}  \right] \frac{1}{x} e^{-\frac{x}{b}} \quad &(x\geq c) 
\end{cases},
\end{split}
\end{equation}
where we defined
\begin{equation}
c=2\mu\alpha r_0.
\end{equation}

\paragraph{Finite Square Well} The finite square well takes the form
\begin{equation}
\label{eq:Vsw}
\begin{split}
V(r)&=
\begin{cases}
-V_0 &(r < r_0)
\\
0 &(r\geq r_0) 
\end{cases}
\\
\tilde{V}(x)&=
\begin{cases}
-\tilde{V}_0 &(x < c)
\\
0 &(x\geq c) ,
\end{cases}
\end{split}
\end{equation}
where we defined 
\begin{equation}
\tilde{V}_0=\frac{V_0}{2 \mu \alpha^2}.
\end{equation}
Notice that for a finite square well $\alpha$ and hence $c$ are not \emph{a priori} defined but are introduced in anticipation of making the connection to the extended Yukawa potential.

\subsection{The Finite Square Well as Limit of Extended Yukawa}
\begin{figure}[H]
\includegraphics[width=1.0\textwidth]{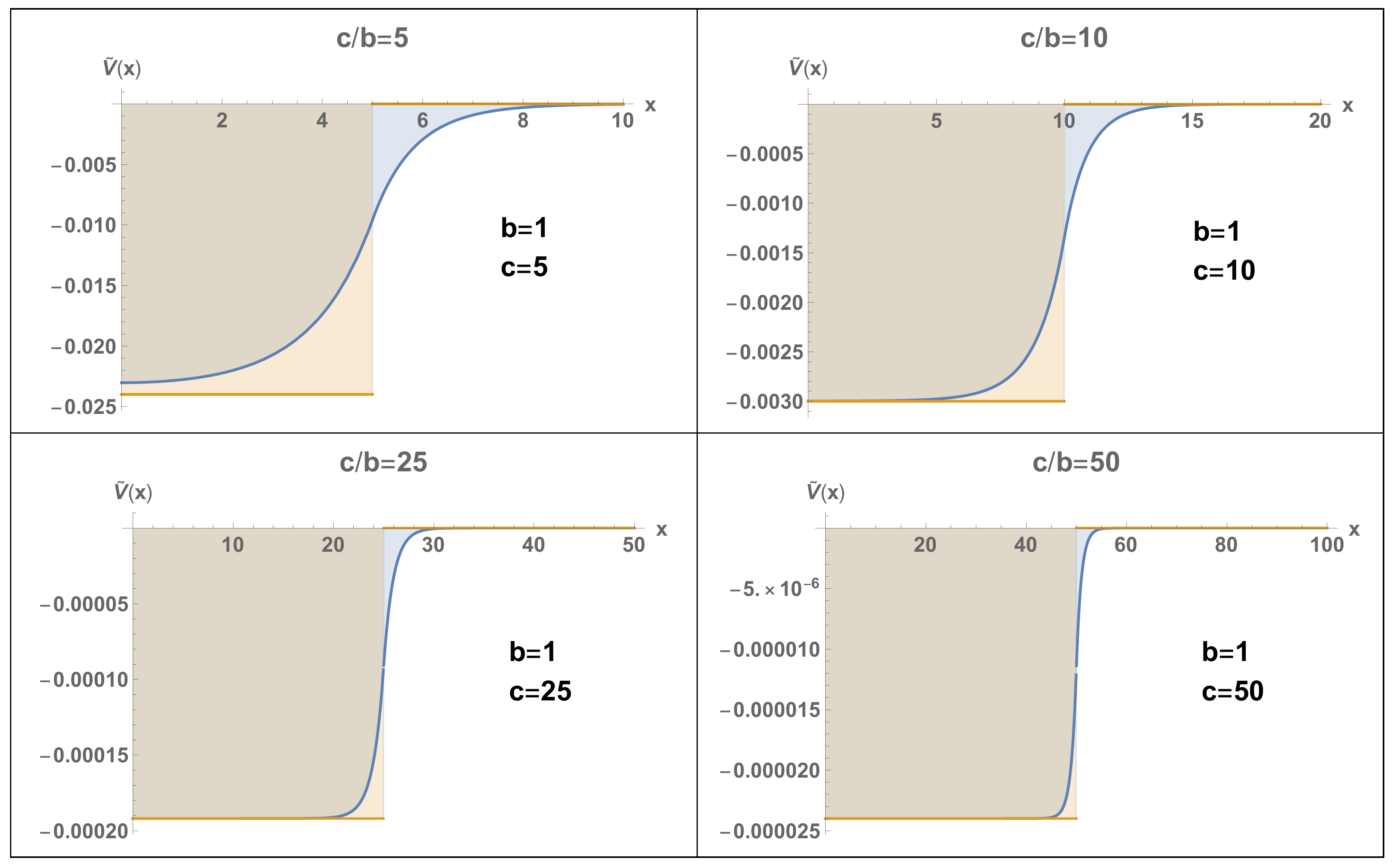}
\caption{ $\tilde{V}(x)$ for extended Yukawa (blue) versus square well (orange) potential for several parameter values.}
\label{fig:ComparePotential}
\end{figure}

Identifying the depth of a finite square well $V_0$ with $V(0)$ of an extended Yukawa,  and keep the radius $r_0$ as a free parameter of either, we have
\begin{equation}
\begin{split}
V_0 &= \frac{3 \alpha}{m_{\phi}^{2} r_0^3}
\\
\tilde{V}_0 &= \frac{3b^2}{c^3}.
\end{split}
\end{equation} 
The finite square well can be thought of as a limit of the extended Yukawa when
\begin{equation}
\begin{split}
\frac{r_0}{r_\phi} &\rightarrow \infty, \quad V_0 = \rm{const}
\\
\frac{c}{b} &\rightarrow \infty, \quad \tilde{V}_0 = \rm{const}.
\end{split}
\end{equation}
Fig.~\ref{fig:ComparePotential} compares both potential for different values of $c/b$.

\subsection{Bound State and Scattering State of Square Well}
The condition for the formation a zero energy bound state is 
\begin{equation}
\label{eq:boundstateconditio}
\tilde{V}_0 c^2 = 2\mu r_0^2 V_0 = (n+\frac{1}{2})^2\pi^2. \quad \quad \quad (n=0,1,2..)
\end{equation}
The scattering state phase shift is determined at $r=r_0$ or $x=c$ by
\begin{equation}
\label{phaseSquareWell}
\begin{split}
&\delta_l=\arctan \left[ {\frac{a c j'_l(a c)-\beta_l j_l (a c)}{a c n'_l(a c)-\beta_l n_l (a c)}} \right],
\\
& \beta_l \equiv \frac{Kc j'_l(Kc)}{j_l(Kc)}
\end{split}
\end{equation}
where $j_l'(Kc) \equiv j_l'(z)$ evaluated at $z=Kc$. And $K = \sqrt{a^2+\tilde{V_0}})$. The following is also true,
\begin{equation}
\label{eq:deltal}
\delta_l=\arctan \left[ {\frac{a  j'_l(a c) j_l(Kc)- K j_l (a c) j_l'(Kc)}{a  n'_l(a c) j_l(Kc)- K n_l (a c) j_l'(Kc)}} \right].
\end{equation}
It is also good to remember that $Kc$ and $ac$ are dimensionless and they can be expressed using dimension-full variables:
\begin{equation}
\begin{split}
ac &=\frac{v}{2 \alpha} \cdot 2\mu\alpha r_0 = \mu v \cdot r_0 = k r_0
\\
Kc &=\sqrt{a^2+\tilde{V}_0} \cdot c = \sqrt{\left(\frac{v}{2\alpha}\right)^2+\frac{V_0}{2\mu \alpha}} \cdot 2 \mu \alpha r_0 =\sqrt{(\mu v)^2 + 2\mu V_0} \cdot r_0 
\\
&=\sqrt{k^2 + 2\mu V_0} \cdot r_0 =\kappa r_0
\end{split}
\end{equation}
where 
\begin{equation}
\kappa=\sqrt{k^2 + 2\mu V_0}=\sqrt{2\mu \left( \frac{k^2}{2\mu} + V_0   \right)}. 
\end{equation}
\subsection{S-wave Resonance and Zero Energy Bound State}
The s-wave cross section is given by
\begin{equation}
\sigma_0=\frac{4\pi}{k^2}\sin^2{\delta_0}=4 \pi r_0^2 \cdot \frac{\sin^2{\delta_0}}{(kr_0)^2} = 4 \pi r_0^2 \cdot \frac{\sin^2{\delta_0}}{(ac)^2}
\end{equation}
and the phase shift can be obtained from Eq.~\eqref{eq:deltal}
\begin{equation}
\label{eq:delta0}
\delta_0 = \arctan{\left[ \frac{ac \cos{(ac)}\sin{(Kc)} - Kc \cos{(Kc)} \sin{(ac)}}{ac \sin{(ac)}\sin{(Kc)} + Kc \cos{(Kc)} \cos{(ac)}} \right]}.
\end{equation}
The cross section is then
\begin{equation}
\label{eq:sigmapQV}
\begin{split}
\frac{\sigma_0}{4\pi r_0^2} &= \frac{1}{(ac)^2}\frac{(Kc \cos{(Kc)} \sin{(ac)} - ac \cos{(ac)}\sin{(Kc)})^2}{ (Kc)^2 \cos^2{(Kc)}+ (ac)^2 \sin^2{(KC)} }
\\
& = \frac{1}{p^2} \frac{(p \cos{p} \sin{Q} - Q \cos{Q}\sin{p})^2}{p^2 \sin^2{Q}+Q^2\cos^2{Q}}
\end{split}
\end{equation}
where we defined
\begin{equation}
\begin{split}
p &\equiv ac = k r_0 
\\
Q &\equiv Kc = \kappa r_0 = \sqrt{p^2 + V^2}
\\
V^2 &\equiv \tilde{V}_{0} c^{2}=2 \mu r_{0}^{2} V_{0}~.
\end{split} 
\end{equation}
For low energy s-wave scattering in the resonant region we have
\begin{equation}
p\ll 1 \leq V \sim Q
\end{equation}
and a zero energy bound state appears when $V=(n+1/2)\pi$.

For small $p$ the cross section goes like
\begin{equation}
\frac{\sigma_0}{4 \pi r_0^2} \sim  \frac{1}{p^2} \frac{p^2(\sin{Q}-Q\cos{Q})^2 + \mathcal{O}(p^4)}{p^2 \sin^2{Q}+Q^2\cos^2{Q}}.
\end{equation}
There are three possibilities in the low energy limit $p \rightarrow  0$:
\paragraph{(1) Constant ($Q\cos{Q} \neq0$  and $\sin{Q}-Q\cos{Q} \neq 0$)} In this case the constant term dominates the denominator so that the cross section goes to a constant:
\begin{equation}
\frac{\sigma_0}{4 \pi r_0^2} \rightarrow  \frac{1}{p^2} \frac{p^2(\sin{Q}-Q\cos{Q})^2 }{Q^2\cos^2{Q}}\rightarrow \frac{(\sin{V}-V\cos{V})^2 }{V^2\cos^2{V}}
\end{equation}

\paragraph{(2) Resonance ($Q\cos{Q} = 0$  and $\sin{Q}-Q\cos{Q} \neq 0$)} In this case the $p^2$ term dominates the denominator, and $\cos{Q}=0$ so that the cross section goes as $p^{-2}$:
\begin{equation}
\frac{\sigma_0}{4 \pi r_0^2} \rightarrow  \frac{1}{p^2} \frac{p^2(\sin{Q}-Q\cos{Q})^2 }{p^2\sin^2{Q}}\sim \frac{1}{p^2}
\end{equation}
Because $p \ll 1$, the cross section can be much larger than the geometrical cross section. 

\paragraph{(3) Anti-Resonance ($Q\cos{Q} \neq 0$  and $\sin{Q}-Q\cos{Q} = 0$)} In this case the $p^2$ term in the numerator vanishes, and $\tan{Q}=Q$. The cross section is suppressed by $p^2$:
\begin{equation}
\frac{\sigma_0}{4 \pi r_0^2} \rightarrow  \frac{1}{p^2} \frac{\mathcal{O}(p^4)}{Q^2\cos^2{Q}}\sim \mathcal{O}(p^2).
\end{equation}
So the p-wave contribution starts to be important. 

\subsection{Near the Resonance}
We expand near the resonance to study the velocity dependence behavior of the s-wave cross section. To be precise, we expand Eq.~\eqref{eq:sigmapQV} around $V=(n+1/2)\pi$:
\begin{equation}
V = \left( n+ \frac{1}{2}\right) \pi + \Delta
\end{equation}
with $\Delta \ll 1$ and $p \ll 1$.  We have:

\begin{equation}
\label{eq:nearRS}
\begin{split}
\frac{\sigma_{0}}{4 \pi r_{0}^{2}}&= \frac{1}{p^2} \frac{(p \cos{p} \sin{Q} - Q \cos{Q}\sin{p})^2}{p^2 \sin^2{Q}+Q^2\cos^2{Q}}
\\
&= \frac{( \cos{\Delta}+ (\frac{2n+1}{2}\pi+\Delta)\sin{\Delta})^2+\mathcal{O}(p^2)}{p^2 (1+(\frac{2n+1}{2}\pi+\Delta)\sin{\Delta}\cos{\Delta}))+(\frac{2n+1}{2}\pi+\Delta)^2\sin^2{\Delta}+\mathcal{O}(p^4)}
\\
&= \frac{1+\mathcal{O}(\Delta,p)}{p^2 +(\frac{2n+1}{2}\pi)^2 {\Delta}^2+\mathcal{O}(\Delta^3 , p^3)}
\\
&\sim  \frac{1}{p^2 +(\frac{2n+1}{2}\pi)^2 {\Delta}^2}. 
\end{split}
\end{equation}
and on resonance $\Delta=0$ while off resonance $\Delta \neq0$
\begin{equation}
\begin{split}
\frac{\sigma_{0}}{4 \pi r_{0}^{2}} &\sim  \frac{1}{(\frac{2n+1}{2}\pi)^2 {\Delta}^2} \quad \quad\quad p\ll \frac{2n+1}{2}\pi \Delta \lesssim 1
\\
\frac{\sigma_{0}}{4 \pi r_{0}^{2}} &\sim  \frac{1}{p^2} \quad  \quad \quad\quad \quad \quad\quad \frac{2n+1}{2}\pi \Delta \ll  p \lesssim 1
\end{split}
\end{equation}

Recalling the definitions:
\begin{equation}
\begin{split}
 p &= ac = k r_0 = \mu v r_0
\\
\Delta &=V-\frac{2n+1}{2}\pi = \sqrt{\tilde{V}_0}c-\frac{2n+1}{2}\pi= \sqrt{2 \mu V_0}r_0-\frac{2n+1}{2}\pi
\\
&=\sqrt{\frac{3b}{c/b}}-\frac{2n+1}{2}\pi=\sqrt{\frac{6\mu \alpha}{m_\phi^2 r_0}}-\frac{2n+1}{2}\pi
\\
\tilde{V}_0&= \frac{V_0}{2 \mu \alpha^2} (\rm{square \, well}) = \frac{3b^2}{c^3} (\rm{Yukawa}),
\end{split}
\end{equation}
and re-expressing Eq.~\eqref{eq:nearRS} in physical variables for the case of the first resonance, we have:
\begin{equation}
\label{eq:sigma_approx}
\sigma  \approx  \frac{4 \pi}{\mu^2 v^2 + \frac{3\mu \pi^2}{2 m_\phi^2 r_0^3}(\sqrt{\alpha} - \sqrt{ \alpha_{\rm res} })^2 } 
\end{equation}
where
\begin{equation}\label{eq:alphares}
 \alpha_{\rm res}  = \left( \frac{\pi}{2} \right)^2 \frac{m_\phi^2 r_0}{6 \mu}.
\end{equation}
Thus
\begin{equation}
\label{eq:hi_v}
			\sigma \rightarrow 4 \pi / (\mu v)^2 ~~~~~~~~~~~~~~~~ v \gg v^* 
\end{equation}
and
\begin{equation}
\label{eq:lo_v}
			\sigma \rightarrow  \frac{8 m_\phi^2 r_0^3}{3 \pi \mu  \left(\sqrt{\alpha} - \sqrt{\alpha_{\rm res}}\right)^2 }~~~~ v \ll v^* ~,
\end{equation}
with
\begin{equation}
\label{eq:v*}
v^* \equiv \frac{\pi |\sqrt{\alpha} - \sqrt{\alpha_{\rm res}}|}{m_\phi r_0\sqrt{2\mu r_0/3}}~.
\end{equation}

\begin{figure}
\centering
\includegraphics[width=0.6\textwidth]{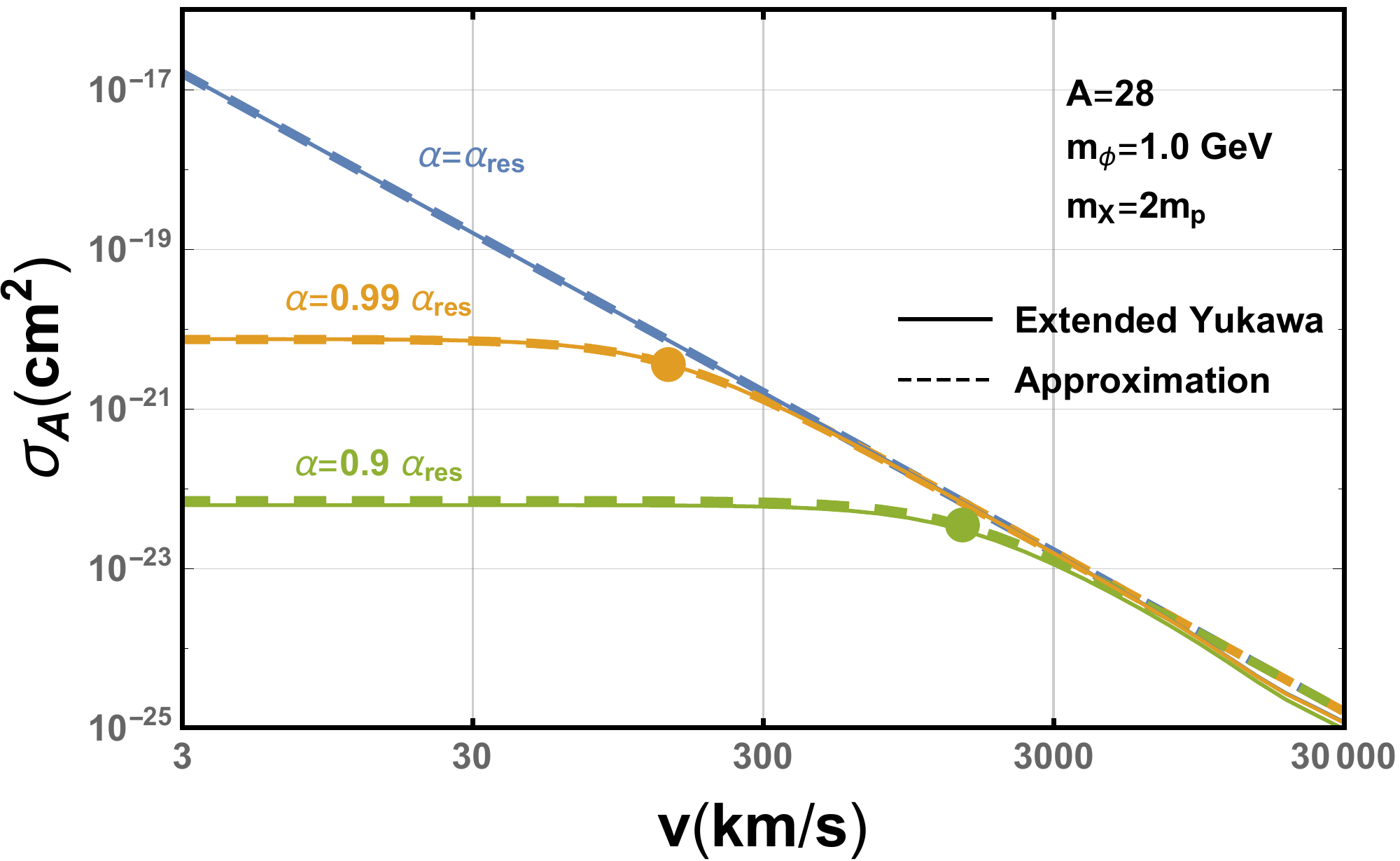}
\caption{$\sigma_{28}(v)$ on and near the first resonance. Solid lines are exact numerical results with and dashed lines are the approximation from Eq.~\eqref{eq:sigma_approx}. The orange and green dots are the transition point ($v^*,\sigma(v^*)$) where $v^*$ is calculated from Eq.~\eqref{eq:v*} and $\sigma(v^*)=2\pi /(\mu v^*)^2 $. Notice there is a slight difference in the location of the resonance: $\alpha_{\rm{res}} =0.129$ for extended Yukawa and $\alpha_{\rm{res}}=0.127$ for the approximation~\ref{eq:alphares}. For A=28 and $m_\phi=1.0 \rm{\,GeV}$, we have $c/b=15.2$. }
\label{fig:vstar}
\end{figure}

Fig.~\ref{fig:vstar} shows how well the approximations Eq.~\eqref{eq:hi_v} - ~\ref{eq:v*} approximate the exact velocity dependence for A=28 (Si) near the first resonance.

\section{Limits for small $m_\phi$}

\begin{figure}
\centering
\includegraphics[width=0.9\textwidth]{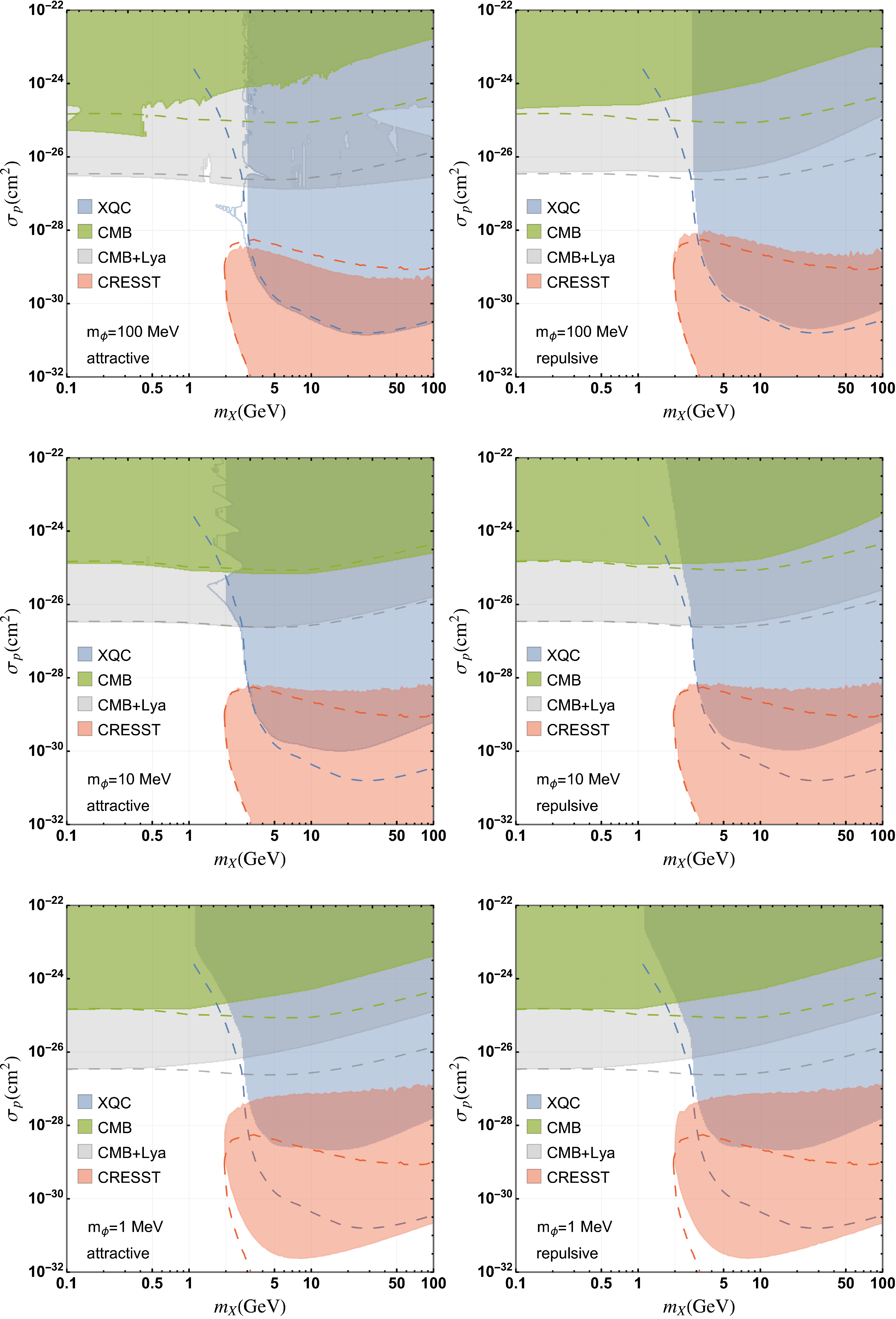}
\caption{\label{fig:smallmphi} Exclusion on ($\sigma_p,m_X$) for $m_\phi=(1,10,100)$ MeV from XQC, CMB (including He-4 scattering) and CRESST. The dashed lines indicate the reported limits assuming Born approximation. The exclusion regions are different from Fig.~\ref{fig:sigmxAll} due to the changed $m_\phi$. For example, when $m_\phi$=1 MeV $\ll 1/r_A$, the cross sections $\sigma_A$ are all in the Born regime for $\sigma_A \lesssim 10^{-21}\rm{\,\,cm}^2$, so the cross section does not depend on the sign of the interaction and there is no (anti-)resonance. However the commonly-used "Born scaling" with $A$   ~\eqref{eq:Ascaling} still does not work, because it relies on the applicability of the low energy condition: $\mu v \ll m_\phi$, which is violated with such small $m_\phi$.}
\end{figure}

\end{document}